\documentclass[twocolumn,multicol]{aastex701}

\begin{document}

\def\la{\mathrel{\hbox{\rlap{\hbox{\lower4pt\hbox{$\sim$}}}\hbox{$<$}}}}
\def\ga{\mathrel{\hbox{\rlap{\hbox{\lower4pt\hbox{$\sim$}}}\hbox{$>$}}}}

\font\sevenrm=cmr7
\def\OIII{[O~{\sevenrm III}]}
\def\FeII{Fe~{\sevenrm II}}
\def\FeIIf{[Fe~{\sevenrm II}]}
\def\SIII{[S~{\sevenrm III}]}
\def\HeI{He~{\sevenrm I}}
\def\HeII{He~{\sevenrm II}}
\def\OI{O~{\sevenrm I}}
\def\CIV{C~{\sevenrm IV}}
\def\NeV{[Ne~{\sevenrm V}]}
\def\OIV{[O~{\sevenrm IV}]}

\def\iraf{{\sevenrm IRAF}}
\def\mpfit{{\sevenrm MPFIT}}
\def\galfit{{\sevenrm GALFIT}}
\def\prepspec{{\sevenrm PrepSpec}}
\def\mapspec{{\sevenrm mapspec}}
\def\cream{{\sevenrm CREAM}}
\def\javelin{{\sevenrm JAVELIN}}
\def\cloudy{{\sevenrm CLOUDY}}
\def\banzai{{\sevenrm BANZAI}}
\def\orac{{\sevenrm ORAC}}
\def\demc{{\sevenrm DEMC}}

\def\gp{\mathcal{GP}}

\title{AGN STORM 2. XI. Spectroscopic reverberation mapping of the hot dust in Mrk~817}

\correspondingauthor{Hermine Landt}

\author[0000-0001-8391-6900]{Hermine Landt}
\affiliation{Centre for Extragalactic Astronomy, Department of Physics, Durham University, South Road, Durham, DH1 3LE, UK}
\email[show]{hermine.landt@durham.ac.uk}

\author[0000-0001-6301-570X]{Benjamin D. Boizelle}
\affiliation{Department of Physics and Astronomy, N283 ESC, Brigham Young University, Provo, UT 84602, USA}
\email{boizellb@byu.edu}

\author[0000-0002-1207-0909]{Michael S. Brotherton}
\affiliation{Department of Physics and Astronomy, University of Wyoming, Laramie, WY 82071, USA}
\email{mbrother@uwyo.edu}

\author[0000-0002-8224-1128]{Laura Ferrarese}
\affiliation{NRC Herzberg Astronomy and Astrophysics Research Centre, 5071 West Saanich Road, Victoria, BC, V9E 2E7, Canada}
\email{Laura.Ferrarese@nrc-cnrc.gc.ca}

\author[0000-0002-3365-8875]{Travis Fischer}
\affiliation{Space Telescope Science Institute, 3700 San Martin Drive, Baltimore, MD 21218, USA}
\email{tfischer@stsci.edu}

\author[0000-0002-8990-2101]{Varoujan Gorjian}
\affiliation{Jet Propulsion Laboratory, M/S 169-327, 4800 Oak Grove Drive, Pasadena, CA 91109, USA}
\email{varoujan.gorjian@jpl.nasa.gov}

\author[0000-0003-0634-8449]{Michael D. Joner}
\affiliation{Department of Physics and Astronomy, N283 ESC, Brigham Young University, Provo, UT 84602, USA}
\email{joner@byu.edu}

\author[0000-0001-8638-3687]{Daniel Kynoch}
\affiliation{School of Physics and Astronomy, University of Southampton, Highfield, Southampton SO17 1BJ, UK}
\email{d.kynoch@soton.ac.uk}

\author[0000-0003-1081-2929]{Jacob N. McLane}
\affiliation{Department of Physics and Astronomy, University of Wyoming, Laramie, WY 82071, USA}
\email{jmclane@uwyo.edu}

\author[0000-0002-5493-1420]{Jake A. J. Mitchell}
\affiliation{Centre for Extragalactic Astronomy, Department of Physics, Durham University, South Road, Durham, DH1 3LE, UK}
\affiliation{Leibniz-Institut f\"{u}r Astrophysik Potsdam (AIP), An der Sternwarte 16, 14482 Potsdam, Germany}
\email{jmitchell@aip.de}

\author[0000-0001-5639-5484]{John W. Montano}
\affiliation{Department of Physics and Astronomy, 4129 Frederick Reines Hall, University of California, Irvine, CA, 92697-4575, USA}
\email{montano3@uci.edu}

\author[0000-0003-0483-3723]{Rogemar A. Riffel}
\affiliation{Departamento de F\'{i}sica, CCNE, Universidade Federal de Santa Maria, Av. Roraima 1000, 97105-900 Santa Maria, RS, Brazil}
\email{rogemar@ufsm.br}

\author[0000-0002-9238-9521]{David Sanmartim}
\affiliation{Rubin Observatory Project Office, 950 N. Cherry Ave., Tucson, AZ 85719, USA}
\email{dsanmartim@lsst.org}

\author[0000-0003-1772-0023]{Thaisa Storchi-Bergmann}
\affiliation{Departamento de Astronomia - IF, Universidade Federal do Rio Grande do Sul, CP 15051, 91501-970 Porto Alegre, RS, Brazil}
\email{thaisa@ufrgs.br}

\author[0000-0003-1810-0889]{Martin J. Ward}
\affiliation{Centre for Extragalactic Astronomy, Department of Physics, Durham University, South Road, Durham, DH1 3LE, UK}
\email{martin.ward@durham.ac.uk}

\author[0000-0002-3026-0562]{Aaron J. Barth}
\affiliation{Department of Physics and Astronomy, 4129 Frederick Reines Hall, University of California, Irvine, CA 92697-4575, USA}
\email{barth@uci.edu}

\author[0000-0002-8294-9281]{Edward M. Cackett}
\affiliation{Department of Physics and Astronomy, Wayne State University, 666 W.\ Hancock St, Detroit, MI 48201, USA}
\email{ecackett@wayne.edu}

\author[0000-0003-3242-7052]{Gisella De~Rosa}
\affiliation{Space Telescope Science Institute, 3700 San Martin Drive, Baltimore, MD 21218, USA}
\email{gderosa@stsci.edu}

\author[0000-0001-8598-1482]{Rick Edelson} 
\affiliation{Eureka Scientific Inc., 2452 Delmer St. Suite 100, Oakland, CA 94602, USA}
\email{rickedelson@gmail.com}

\author[0000-0001-9092-8619]{Jonathan Gelbord}
\affiliation{Spectral Sciences Inc., 30 Fourth Ave., Suite 2, Burlington, MA 01803, USA}
\email{jgelbord@gmail.com}

\author[0000-0002-0957-7151]{Yasaman Homayouni}
\affiliation{Space Telescope Science Institute, 3700 San Martin Drive, Baltimore, MD 21218, USA}
\affiliation{Department of Astronomy and Astrophysics, The Pennsylvania State University, 525 Davey Laboratory, University Park, PA 16802, USA}
\affiliation{Institute for Gravitation and the Cosmos, The Pennsylvania State University, University Park, PA 16802, USA}
\email{ybh5251@psu.edu}

\author[0000-0003-1728-0304]{Keith Horne}
\affiliation{SUPA Physics and Astronomy, University of St. Andrews, Fife, KY16 9SS, UK}
\email{kdh1@st-andrews.ac.uk}

\author[0000-0003-0172-0854]{Erin A. Kara}
\affiliation{MIT Kavli Institute for Astrophysics and Space Research, Massachusetts Institute of Technology, 77 Massachusetts Avenue, Cambridge, MA 02139, USA}
\email{ekara@mit.edu}

\author[0000-0002-2180-8266]{Gerard A. Kriss}
\affiliation{Space Telescope Science Institute, 3700 San Martin Drive, Baltimore, MD 21218, USA}
\email{gak@stsci.edu}

\author[0000-0003-2991-4618]{Nahum Arav}
\affiliation{Department of Physics, Virginia Tech, Blacksburg, VA 24061, USA}
\email{arav@vt.edu}

\author[0000-0001-9931-8681]{Elena Dalla Bont\`{a}}
\affiliation{Dipartimento di Fisica e Astronomia ``G. Galilei,'' Universit\'{a} di Padova, Vicolo dell'Osservatorio 3, I-35122 Padova, Italy}
\affiliation{INAF-Osservatorio Astronomico di Padova, Vicolo dell'Osservatorio 5 I-35122, Padova, Italy}
\affiliation{Jeremiah Horrocks Institute, University of Central Lancashire, Preston, PR1 2HE, UK}
\email{elena.dallabonta@unipd.it}

\author[0000-0002-0964-7500]{Maryam Dehghanian}
\affiliation{Department of Physics and Astronomy, The University of Kentucky, Lexington, KY 40506, USA}
\email{m.dehghanian@uky.edu}

\author[0000-0003-4503-6333]{Gary J. Ferland}
\affiliation{Department of Physics and Astronomy, The University of Kentucky, Lexington, KY 40506, USA}
\email{gary@g.uky.edu}

\author[0000-0002-2306-9372]{Carina Fian}
\affiliation{Departamento de Astronom\'{i}a y Astrof\'isica, Universidad de Valencia, E-46100 Burjassot, Valencia, Spain}
\affiliation{Observatorio Astron\'{o}mico, Universidad de Valencia, E-46980 Paterna, Valencia, Spain}
\email{carina.fian@uv.es}

\author[0000-0002-9280-1184]{Diego H. Gonz\'{a}lez Buitrago}
\affiliation{Instituto de Astronom\'{\i}a, Universidad Nacional Aut\'{o}noma de M\'{e}xico, Km 103 Carretera Tijuana-Ensenada, 22860 Ensenada B.C., M\'{e}xico}
\email{dgonzalez@astro.unam.mx}

\author[0000-0002-1134-4015]{Dragana Ili\'{c}}
\affiliation{University of Belgrade - Faculty of Mathematics, Department of Astronomy, Studentski trg 16, 11000 Belgrade, Serbia}
\affiliation{Hamburger Sternwarte, Universit\"at Hamburg, Gojenbergsweg 112, 21029 Hamburg, Germany}
\email{dragana.ilic@matf.bg.ac.rs}

\author[0000-0002-9925-534X]{Shai Kaspi}
\affiliation{School of Physics and Astronomy and Wise Observatory, Tel-Aviv University, Tel-Aviv 6997801, Israel}
\email{shaik@tauex.tau.ac.il}

\author[0000-0001-6017-2961]{Christopher S. Kochanek}
\affiliation{Department of Astronomy, The Ohio State University, 140 West 18th Avenue, Columbus, OH 43210, USA}
\affiliation{Center for Cosmology and Astroparticle Physics, The Ohio State University, 191 West Woodruff Avenue, Columbus, OH 43210, USA}
\email{kochanek.1@osu.edu}

\author[0000-0001-5139-1978]{Andjelka B. Kova\v{c}evi\'c}
\affiliation{University of Belgrade - Faculty of Mathematics, Department of Astronomy, Studentski trg 16, 11000 Belgrade, Serbia}
\email{andjelka.kovacevic@matf.bg.ac.rs}

\author[0000-0002-8671-1190]{Collin Lewin}
\affiliation{MIT Kavli Institute for Astrophysics and Space Research, Massachusetts Institute of Technology, 77 Massachusetts Avenue, Cambridge, MA 02139, USA}
\email{clewin@mit.edu}

\author[0000-0001-5841-9179]{Yan-Rong Li}
\affiliation{Key Laboratory for Particle Astrophysics, Institute of High Energy Physics, Chinese Academy of Sciences, 19B Yuquan Road, Beijing 100049, People's Republic of China}
\email{liyanrong@mail.ihep.ac.cn}

\author[0000-0002-4994-4664]{Missagh Mehdipour}
\affiliation{Space Telescope Science Institute, 3700 San Martin Drive, Baltimore, MD 21218, USA}
\email{mmehdipour@stsci.edu}

\author[0000-0002-6766-0260]{Hagai Netzer}
\affiliation{School of Physics and Astronomy, Tel-Aviv University, Tel-Aviv 6997801, Israel}
\email{hagainetzer@gmail.com}

\author[0000-0002-2509-3878]{Rachel Plesha}
\affiliation{Space Telescope Science Institute, 3700 San Martin Drive, Baltimore, MD 21218, USA}
\email{rplesha@stsci.edu}

\author[0000-0003-2398-7664]{Luka \v{C}. Popovi\'{c}}
\affiliation{Astronomical Observatory Belgrade, Volgina 7, 11000 Belgrade, Serbia}
\affiliation{University of Belgrade - Faculty of Mathematics, Department of Astronomy, Studentski trg 16, 11000 Belgrade, Serbia}
\email{lpopovic@aob.rs}

\author[0000-0002-6336-5125]{Daniel Proga}
\affiliation{Department of Physics \& Astronomy, University of Nevada, Las Vegas, 4505 S.\ Maryland Pkwy, Las Vegas, NV, 89154-4002, USA}
\email{daniel.proga@unlv.edu}

\author[0000-0001-9449-9268]{Jian-Min Wang}
\affiliation{Key Laboratory for Particle Astrophysics, Institute of High Energy Physics, Chinese Academy of Sciences, 19B Yuquan Road, Beijing 100049, People's Republic of China}
\affiliation{School of Astronomy and Space Sciences, University of Chinese Academy of Sciences, 19A Yuquan Road, Beijing 100049, People's Republic of China}
\affiliation{National Astronomical Observatories of China, 20A Datun Road, Beijing 100020, People's Republic of China}
\email{wangjm@ihep.ac.cn}

\author[0000-0003-0931-0868]{Fatima Zaidouni}
\affiliation{MIT Kavli Institute for Astrophysics and Space Research, Massachusetts Institute of Technology, 77 Massachusetts Avenue, Cambridge, MA 02139, USA}
\email{fzaid@mit.edu}

\author[0000-0001-6966-6925]{Ying Zu}
\affiliation{Department of Astronomy, School of Physics and Astronomy, Shanghai Jiao Tong University, 800 Dongchuan Road, Shanghai, 200240, People's Republic of China}
\affiliation{Shanghai Key Laboratory for Particle Physics and Cosmology, Shanghai Jiao Tong University, Shanghai 200240, People's Republic of China}
\email{yingzu@sjtu.edu.cn}

\begin{abstract}

The AGN Space Telescope and Optical Reverberation Mapping 2 (STORM~2) campaign targeted Mrk~817 with intensive multi-wavelength monitoring and found its soft X-ray emission to be strongly absorbed. We present results from 157 near-IR spectra with an average cadence of a few days. Whereas the hot dust reverberation signal as tracked by the continuum flux does not have a clear response, we recover a dust reverberation radius of $\sim 90$~light-days from the blackbody dust temperature light-curve. This radius is consistent with previous photometric reverberation mapping results when Mrk~817 was in an unobscured state. The heating/cooling process we observe indicates that the inner limit of the dusty torus is set by a process other than sublimation, rendering it a luminosity-invariant `dusty wall' of a carbonaceous composition. Assuming thermal equilibrium for dust optically thick to the incident radiation, we derive a luminosity of $\sim 6 \times 10^{44}$~erg~s$^{-1}$ for the source heating it. This luminosity is similar to that of the obscured spectral energy distribution, assuming a disk with an Eddington accretion rate of $\dot{m} \sim 0.2$. Alternatively, the dust is illuminated by an unobscured lower luminosity disk with $\dot{m} \sim 0.1$, which permits the UV/optical continuum lags in the high-obscuration state to be dominated by diffuse emission from the broad-line region. Finally, we find hot dust extended on scales $\ga 140-350$~pc, associated with the rotating disk of ionised gas we observe in spatially-resolved \SIII~$\lambda 9531$ images. Its likely origin is in the compact bulge of the barred spiral host galaxy, where it is heated by a nuclear starburst.
\end{abstract}


\section{Introduction}

Infrared (IR) emission stretching from near-IR to far-IR wavelengths is a prominent part of the multi-wavelength spectral energy distributions (SEDs) of active galactic nuclei (AGN) and it is usually attributed to thermal radiation from dust. The nuclear dusty structure is commonly assumed to be optically thick to the ultraviolet (UV)/optical radiation heating it and to have a toroidal geometry with the torus aligned with the plane of the accretion disk, extending over a few pc. This geometry is needed for an equatorial obscurer in order to explain the broad emission lines in the polarized, scattered light seen in many type~2 AGN and the relative numbers of type~1 and type~2 AGN \citep[see reviews by][]{Law87, Ant93, Netzer15, Lyu22a}. A warped, dusty disk could be a viable alternative to the dusty torus, as first proposed by \citet{Phi89b}. The extended dusty torus is expected to have a radial temperature structure, irrespective of whether it is continuous or clumpy, with the hottest, near-IR emitting dust located closest to the central supermassive black hole and warm, mid-IR emitting dust (of a few 100~K) located further out \citep{Weed05, Buch06, L10b, Lyu18, Brown19}. Cold dust (of a few 10~K) emitting at far-IR wavelengths in AGN is most likely heated by young stars in the host galaxy rather than by the nuclear UV/optical emission from the accretion disk \citep{Kirk15}.

AGN are ideal laboratories to investigate the chemical composition and grain properties of astrophysical dust. Their UV/optical luminosities are usually high enough to heat the innermost circumnuclear dust to sublimation temperatures. If these highest temperatures can be observed, they can in principle constrain the chemistry since different species condense out of the gas phase in different environmental conditions. Dust temperatures were measured with simultaneous photometry at several near-IR wavelengths in a handful of sources \citep{Clavel89, Glass04, Schnuelle13, Schnuelle15}, but such studies have only come of age with the availability of efficient cross-dispersed near-IR spectrographs. \citet{L11a} and \citet{L14} derived dust temperatures from such spectroscopy for the largest sample of type~1 AGN so far ($\sim 30$ sources). Their measurements yielded a very narrow temperature distribution, with an average value of $T \sim 1400$~K. This means that the hot dust is composed of only silicate dust grains, if it is at its sublimation temperature \citep[$T_{\rm sub} \sim 1300 - 1500$~K;][]{Lod03}, and so formed in an oxygen-rich environment. However, if also carbon dust is present, then it is {\it not} heated close to its sublimation point, since carbonaceous dust (e.g., graphite) can survive up to $T_{\rm sub} \sim 1800 - 2000$~K, with the sublimation temperature slightly dependent on gas pressure \citep{Sal77}.

The range in temperatures within the dusty torus translates to a range in radius and only the mid-IR emitting dust can be spatially resolved with current instruments. High-angular-resolution mid-IR imaging \citep{Ramos11} and mid-IR interferometry \citep[e.g.,][]{Tri09, Bur13} of a dozen nearby and bright AGN delivered useful upper limits on the extent of the dusty torus of at most few parsec. Such spatially resolved observations have also revealed a significant warm dust component perpendicular to the plane of the accretion disk, referred to as 'polar dust' \citep{Tri14, Isbell22, Lyu22b, Gamez22}, which was also evident from SED studies \citep{L10b, Isbell21}. GRAVITY, the near-IR interferometric instrument at the Very Large Telescope (VLT), has recently started to resolve the innermost, hottest part of the central dusty structure in some nearby, luminous sources \citep{gravity20, gravity24} and further progress is expected from an upgrade in sensitivity to GRAVITY$+$. But since most AGN will remain unresolved by near-IR interferometry, in particular low-luminosity sources, knowledge about the location of the inner dust radius is most efficiently obtained for larger samples through dust reverberation, measuring the time lag of the response of the dust to variations in the irradiating accretion disk flux. 

For $\sim 60$ AGN, the hot dust radius was determined by photometric campaigns, often coordinated between optical and near-IR wavelengths \citep[e.g.,][]{Clavel89, Nel96, Okn01, Glass04, Sug06, Schnuelle13, Kosh14, Schnuelle15, Vazquez15, Min19, Lyu19}. In general, observed dust response times follow a luminosity-radius relationship with a slope similar to that for the broad emission line region (BLR), indicating a narrow hot dust temperature distribution. However, dust radii measured via reverberation are often smaller (by factors of a few) than dust radii measured by interferometry or estimated from the SED assuming thermal equilibrium \citep{Kish07, Nen08a, L14}. A possible interpretation for this finding is that the dust has a bowl-shaped geometry caused by the anisotropy of the accretion disk emission irradiating it \citep{Kaw10, Kaw11}. In such a geometry, the hot dust located in the plane of the accretion disk is located closer to the central source than the bulk of the dust and can dominate the reverberation signal because its response is less smoothed in time.

Progress in technology and flexible scheduling have now made spectroscopic near-IR monitoring campaigns feasible. \citet{L19} presented the first such program. They chose the source NGC~5548 since it had been extensively monitored by the AGN Space Telescope and Optical Reverberation Mapping (STORM) program \citep{Storm1} only a couple of years previously. Using near-IR cross-dispersed spectra that provide a wide wavelength coverage extending partially into the optical, they showed that such a campaign could (i) monitor a large portion of the hot dust SED to measure the dust temperature and its evolution; (ii) separate the accretion disk emission from that of the hot dust, which is a considerable source of systematic error in photometric campaigns; (iii) with the measurement of the dust temperature determine the luminosity-weighted dust radius simultaneously with the response-weighted dust radius; (iv) construct the variable near-IR spectrum; and (v) study the variability of emission lines formed in the BLR and the coronal line region \citep{Kynoch22}. Their study of the hot dust in NGC~5548 found that a single component dominated both the mean emission and the variations, with the dust reponse time and the luminosity-based dust radius being consistent with each other only if a blackbody emissivity was assumed. This result constrained the dust grain size to a few $\mu$m. 

The temperature and its variability indicated carbonaceous dust well below the sublimation threshold undergoing a heating and cooling process in response to the variable UV/optical accretion disk flux irradiating it. Most importantly, the dust reverberation signal showed tentative evidence for a second hot dust component most likely associated with the accretion disk. The existence of such dust in the UV radiation field of the disk, which will generally prevent dust formation, is a prerequisite for the recent models of the AGN structure proposed by \citet{Czerny17} and \citet{Baskin18}. These models explain both the BLR and dusty torus as part of the same outflow launched from the outer regions of the accretion disk by radiation pressure on dust, preferentially on carbonaceous dust since it has a higher opacity than silicate dust.

When AGN STORM~2 was conceived, it was decided to incorporate near-IR monitoring from the beginning, allowing the campaign to map the accretion disk simultaneously with the dusty torus and the disk itself from the UV through to the near-IR. The near-IR monitoring would also add a plethora of lines from the BLR, mainly from the Paschen and Brackett hydrogen series and other low-ionisation lines, as well as several high-ionisation coronal lines \citep{L08a}. Here, we present the results on the hot dust in Mrk~817 from the STORM~2 near-IR spectroscopic monitoring campaign, which found this AGN to be affected by strong soft X-ray obscuration, just like NGC~5548 was during STORM. 

This paper differs from others in this series because it also contains data gathered after the end of STORM~2 (2022 Feb 24) by a different program, namely, the Extended Mrk~817 Reverberation Mapping Campaign (the `Extended Campaign'), which initially had a different main goal: to perform Intensive Broadband Reverberation Mapping \citep[IBRM; for details of this technique see][]{Edelson24} on what would become the longest IBRM dataset ever obtained, spanning $\sim 1200$ days. This full data set is being used to characterize the time evolution of the central engine and obscurer over unprecedentedly long timescales (Edelson et al., in prep.).

The structure of our paper is as follows. After we briefly discuss our science target in Section~2, we give details of the observations, data reduction and measurements in Section~3. In Section~4, we study the variability properties of the hot dust for our campaign. We discuss our main results and put them into perspective in Section~5, with a focus on the geometry of the hot dust and the role of the {\it extended} hot dust component detected in this work. Finally, in Section~6, we present a short summary and our conclusions.

\section{The science target} \label{target}

Mrk~817 (PG~1434$+$590) is at a redshift of \mbox{$z=0.03145$} at the center of a barred Sbc spiral galaxy \citep{Bentz09, Bentz18}. It has an average $V$-band luminosity of $\sim 5 \times 10^{43}$~erg~s$^{-1}$ \citep{Bentz13}, which places it at the undersampled bottom end of the relationship between the hot dust radius and optical continuum luminosity presented by \citet{Kosh14} and \citet{Min19}. \citet{Kosh14} carried out a co-ordinated optical and near-IR photometric monitoring campaign during the years \mbox{2003-2006} and measured a $K$ band dust response time of $\tau = 92\pm9$~days.

Mrk~817 has a black hole mass well-determined by optical reverberation mapping campaigns of \mbox{$M_{\rm BH} = (3.9 \pm 0.6) \times 10^7~M_\odot$} \citep{Bentz15}, using a virial product scaling factor of $f=4.3$ \citep{Grier13b}. For this black hole mass, the corresponding gravitational radius is $r_{\rm g} = G M_{\rm BH}/c^2 = 5.8 \times 10^{12}$~cm = 0.002~light-days, with $G$ the gravitational constant and $c$ the speed of light. The corresponding Eddington luminosity is $L_{\rm edd} = 4.9 \times 10^{45}$~erg~s$^{-1}$. 

Mrk~817 (J2000 \mbox{R.A. $14^h 36^m 22.1^s$}, \mbox{Decl. $+58^\circ 47\arcmin 39\arcsec$}) is observable mainly from the northern hemisphere. Its low redshift and high intrinsic luminosity make it sufficiently bright in the near-IR \citep[2MASS $J=12.9$~mag, $K_s=10.9$~mag;][]{2MASS} to obtain a high-quality spectrum in a relatively short exposure time. Its emission-line spectrum is of the inflected type, where the broad line profiles have clearly discernible broad- and narrow-line components. 

We adopt here the cosmological parameters \mbox{$H_0 = 70$~km~s$^{-1}$~Mpc$^{-1}$}, $\Omega_{\rm M}=0.3$, and $\Omega_{\Lambda}=0.7$, which give a luminosity distance to Mrk~817 of 138.8~Mpc and an angular scale at the source of 632~pc/$\arcsec$.

\subsection{Main results from the AGN STORM~2 campaign}

We summarise in the following some relevant findings from the published STORM~2 papers, which studied the properties of Mrk~817 in the obscured state it was found to be in when the campaign started:

\begin{itemize}

\item 
In the X-ray band, three types of ionized winds have been observed so far in AGN: warm absorbers (WAs), ultrafast outflows (UFOs), and obscurers. Mrk~817 harbours an obscurer detected at the beginning of the STORM~2 campaign \citep{Storm2-paper1} and also a UFO detected by \citet{Miller21} and \citet{Zak24}. \citet{Storm2-paper9} determined that the obscurer is a multiphase ionized wind with an outflow velocity of $\sim 5200$~km~s$^{-1}$ emerging at a radius of $\sim 2$~light-days from the central X-ray source. The lower ionization component could be associated with the observed UV absorption features, indicating that the obscurer consists of dense clumps embedded in diffuse, higher-ionization gas. \citet{Storm2-paper9} found no evidence of a WA, which makes Mrk~817 the first example of an AGN with an obscurer but without a WA. 

\item 
An X-ray flare occurred on 2021 Apr 18, which separated a period of high obscuration (pre-flare) from a period of low obscuration (post-flare) \citep{Storm2-paper3}. After $\sim 5$~months, the high obscuration returned and remained. 

\item 
\citet{Storm2-paper1} derived the bolometric luminosities for the unobscured and obscured SEDs as $1.1 \times 10^{45}$~erg~s$^{-1}$ and $6.7 \times 10^{44}$~erg~s$^{-1}$, respectively, for which the corresponding $1–1000$~Ryd (13.6 eV to 13.6 keV) ionizing luminosities are $6.3 \times 10^{44}$~erg~s$^{-1}$ and $2.4 \times 10^{44}$~erg~s$^{-1}$, respectively. The unobscured SED corresponds to an Eddington ratio of $\dot{m} = L/L_{\rm edd} = 0.2$.

\item 
\citet{Storm2-paper1} derived the hydrogen BLR radius from the first third of the optical spectroscopic reverberation mapping campaign to be $23 \pm 2$~light-days (in the rest-frame), based on the response time of the optical Balmer line H$\beta$.

\item 
\citet{Storm2-paper6} assumed that the UV/optical continuum variability is due to accretion disk temperature fluctuations and produced maps for these fluctuations resolved in time and radius. They found evidence for coherent radial structures that move slowly both inward and outward, which conflicts with the idea that disk variability is driven only by reverberation. Instead, these slow-moving temperature fluctuations, which mostly exist over relatively long timescales (hundreds of days), are likely due to variability intrinsic to the disk.

\item
The UV/optical differential lags increase with wavelength approximately following $\tau \propto \lambda^{4/3}$, as expected for a standard geometrically thin, optically thick accretion disk \citep{Storm2-paper1, Storm2-paper4}. Considering frequency-resolved lags, \citet{Storm2-paper7} fit \mbox{$\tau = \tau_0 [(\lambda/\lambda_0)^{4/3} -1 ]$}, with $\lambda_0 = 1869$~\AA~the rest-frame wavelength of the {\it Swift} UVW2 band, and obtained best-fitting values of $\tau_0 = 0.58 \pm 0.03$~days and \mbox{$0.23 \pm 0.01$}~days for timescales of 20-50 days and 9-20 days, respectively, and $\tau_0 = 0.69 \pm 0.03$~days and \mbox{$0.35 \pm 0.02$}~days for periods of high and low obscuration, respectively. The shortest $\tau_0$ values yield lags close to theoretical accretion disk predictions for $\dot{m} = L/L_{\rm edd} = 0.2$. 

\item
\citet{Storm2-paper10} showed that the long UV/optical differential lags measured by \citet{Storm2-paper7} during the periods of high obscuration are consistent with being dominated by emission from the BLR and its associated diffuse continuum (DC) emission. They also presented an alternative explanation for the short lags during the period of low obscuration; rather than being dominated by the accretion disk emission, the lags are shortened due to additional DC emission from the obscurer, which at the same time partly shields the BLR and thus diminishes the BLR DC emission contribution.

\end{itemize}

\section{The observations}

\subsection{The near-IR spectroscopy} \label{spectroscopy}

\begin{deluxetable*}{lllcccccl}
\tablecolumns{9}
\decimalcolnumbers
\tablecaption{\label{campaign} 
General properties of the near-IR spectroscopic campaign}
\tablehead{
\colhead{Telescope} & \colhead{instrument} & \colhead{slit} & \colhead{wavelength coverage} & \colhead{resolution} & \colhead{N$_{\rm spec}$} & \multicolumn{2}{c}{continuum $S/N$} & \colhead{observing periods} \\
&& \colhead{(arcsec$^2$)} & \colhead{(\AA)} &&& \colhead{$H$} & \colhead{$K$} & 
}
\startdata
ARC 3.5~m        & TripleSpec & $1.1 \times 43$ & $9500 - 24600$ & 3500 & 56 &  63 & 102 & 2020 Dec 3 - 2021 Sep 18 \\
Gemini North 8~m & GNIRS      & $0.45 \times 7$ & $8200 - 25190$ & 1100 & 60 & 132 & 154 & 2021 Jan 14 - 2021 Aug 2 \\
                 &            &                 &                &      &    &     &     & 2022 Feb 3 - 2022 Aug 2 \\
IRTF 3~m         & SpeX       & $0.3 \times 15$ & $7000 - 25500$ & 2000 & 41 &  96 & 132 & 2021 Feb 12 - 2021 Sep 4 \\
                 &            &                 &                &      &    &     &     & 2022 Feb 3 - 2022 Jun 22 \\
\enddata

\tablecomments{The columns are: (1) Telescope; (2) cross-dispersed near-IR spectrograph; (3) slit size; (4) wavelength coverage; (5) spectral resolving power $R=\lambda/\Delta\lambda$; (6) number of spectral epochs; average continuum $S/N$ in the rest-frame wavelength regions of (7) $\lambda=1.50-1.55~\mu$m ($H$ band) and (8) $\lambda=2.05-2.10~\mu$m ($K$ band), and (9) observing periods.}

\end{deluxetable*}

\begin{figure*} 
\centerline{
\includegraphics[scale=0.95]{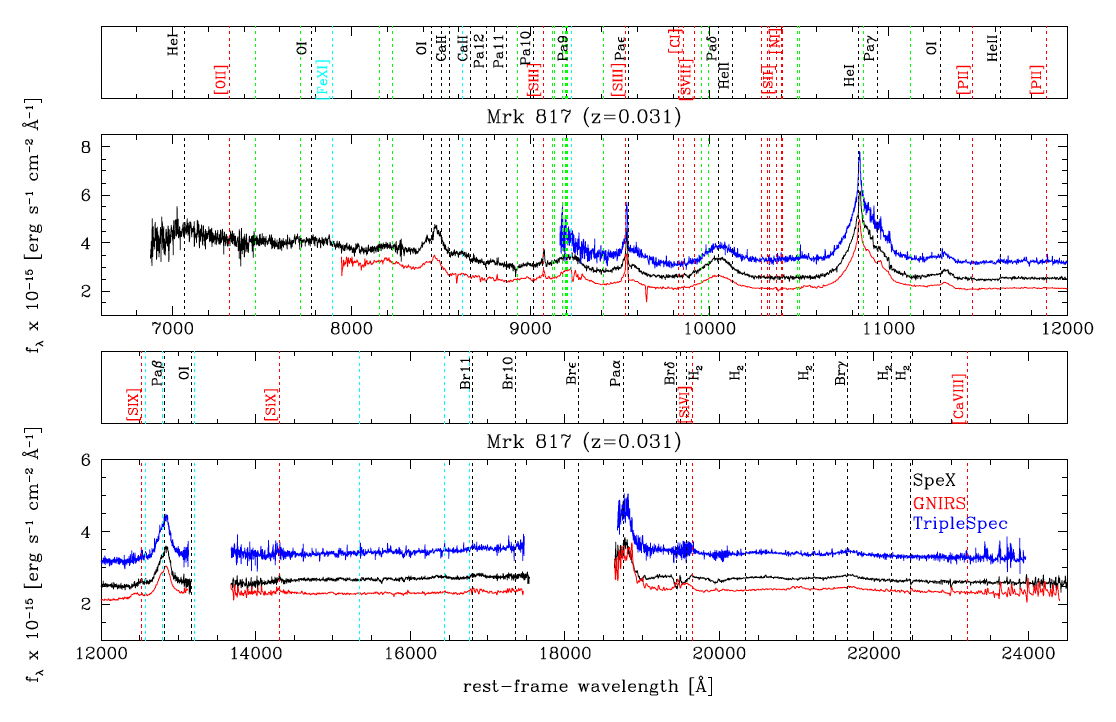}
}
\caption{\label{irspec} Near-IR spectra of Mrk~817 obtained with the cross-dispersed instruments TripleSpec on ARC (on 2020 Dec 26; blue solid line), GNIRS on Gemini North (on 2021 Jan 14; red solid line) and SpeX on IRTF (on 2021 Mar 19; black solid line) shown as observed flux versus rest-frame wavelength. Emission lines listed in Table 4 of \citet{L08a} are marked by dotted lines and labeled; black: permitted transitions, green: permitted \FeII~multiplets (not labeled), red: forbidden transitions and cyan: forbidden transitions of iron (those of \FeIIf~not labeled).}
\end{figure*}

All STORM~2 near-IR spectroscopy was obtained in the cross-dispersed mode, providing a broad wavelength coverage and simultaneously covering the $z$, $J$, $H$ and $K$ bands. We started the spectroscopic coverage on 2020 Dec 3 and continued until the source set in 2021 Sep. We obtained a total of 122 spectra with an average cadence of $\sim 3$~days. We resumed observing Mrk~817 on 2022 Feb 3 and covered a second season until 2022 Aug 2 with a total of 35 spectra at an average cadence of $\sim 6$~days. This remarkable achievement of obtaining a total of 157 spectra over a period of 608 days was only possible by employing three different telescopes. Since Mrk~817 is first accessible at higher latitudes, we started the monitoring with the Astrophysical Research Consortium (ARC)~3.5~m at the Apache Point Observatory, New Mexico, and then added the Gemini North~8~m and NASA Infrared Telescope Facility (IRTF)~3~m, Hawaii, about one and two months later, respectively. Table~\ref{campaign} summarises our near-IR spectroscopic observing campaign and Table \ref{obslog} lists the journal of observations. We show examples of the near-IR spectra from the three telescopes in Fig.~\ref{irspec}. Based on the Galactic hydrogen column densities \citep{DL90}, the Galactic extinction towards Mrk~817 is negligible and we did not correct the spectra for it.

In the following we give details of the observations and data reduction for the three telescopes, which in principle were similar. The important points for the reliability of the spectral light-curves and shapes we derive later are: (i) the large wavelength range of the spectrum is obtained {\it simultaneously} by dispersing 6-7 orders over the detector, which are then stitched together by using a considerable wavelength range of overlap at their respective ends; and (ii) the telluric standard star used for the flux calibration is observed with the same set-up and {\it close} in time and sky position to Mrk~817. The first point means that an absolute flux calibration performed at a given wavelength translates to the entire spectrum (see Section~\ref{fluxscale}). The second point means that the wavelength-dependent variations in the seeing and so in the Point Spread Function (PSF), which could influence the spectral shape (see Section~\ref{dustspectrum}), are well corrected for. In any case, the main influence of clouds in the near-IR is not extinction but an increase in humidity, which decreases the signal in the numerous telluric absorption bands and thus can make the recovery of the signal in these spectral regions difficult.

\subsubsection{ARC~3.5~m} \label{arc}

We used the TripleSpec spectrograph \citep{Wilson04} on the ARC~3.5~m during one observing season. Due to weather and technical issues, 38 of the scheduled observing windows were lost, resulting in successful observations on 56 nights during the period of 2020 Dec 3 to 2021 Sep 18. The seeing during the observations varied over the range of $1\farcs0-1\farcs6$ and was on average $1\farcs2$. TripleSpec is a cross-dispersed spectrograph that covers the \mbox{$0.95 - 2.46~\mu$m} wavelength region with a fixed $1\farcs1 \times 43\arcsec$ slit. In this mode, TripleSpec has an average spectral resolution of $R \sim 3500$ or full width at half-maximum (FWHM) $\sim 85$~km~s$^{-1}$. Due to increased pattern noise when the detector was rotated, we kept the slit along the horizon direction ($+90^\circ$) for the duration of the program. Depending on the weather and allocated time, on-source integrations varied between $4 \times 300$~s and $12 \times 300$~s, with exposures taken in an ABBA pattern nodding along the slit since the source is not significantly extended in the near-IR.

Directly before the science target, we used the same spectral set-up to observe the telluric standard A0~V star HD~121409. Flats were usually obtained after observing the science target. We reduced the data using the Interactive Data Language (IDL) package Triplespectool, which is a modified version of Spextool \citep{Cush04}. Similar to Spextool (see Section~\ref{irtf}), Triplespectool performs the flat-field correction, wavelength calibration, spectral extraction, telluric correction, and flux calibration to produce final, merged spectra. We used the optimally weighted extraction method \citep{Horne86a} to extract the spectra, which accounts for the spatial profile in each order and fits and subtracts a local background. Wavelength calibration was carried out using sky lines in each standard and science frame, with the accuracy increasing from $\sim 0.7$~\AA~in the $J$ band to $\sim 0.3$~\AA~in the $K$ band.

\subsubsection{Gemini North~8~m}

We monitored Mrk~817 at the Gemini North 8~m in queue mode during two observing seasons covering the periods of 2021 Jan 14 to 2021 Aug 2 and 2022 Feb 3 to 2022 Aug 2 (Program ID: \mbox{GN-2021A-Q-124}, \mbox{GN-2021A-Q-215}, \mbox{GN-2022A-Q-109}), with a total of 60 successful epochs. We used the Gemini Near-Infrared Spectrograph \citep[GNIRS;][]{gnirs} in the cross-dispersed mode with the short camera at the lowest spectral resolution (31.7~l~mm$^{-1}$ grating), covering the entire wavelength range of \mbox{$0.85-2.5$~$\mu$m} without inter-order contamination. We chose the $0\farcs45 \times 7\arcsec$ slit, which we oriented at the parallactic angle. This set-up gives an average spectral resolution of $R=1100$ (FWHM $\sim 270$~km~s$^{-1}$), which is sufficient to study line profiles and to clearly separate the narrow and broad line components for the permitted transitions. The narrow slit also minimizes the flux contamination from the host galaxy. The on-source exposure time was $8 \times 120$~s, with individual frames obtained by nodding along the slit.

After or before the science target, we observed the nearby (in position and airmass) stars HIP~78017 (HD~143187) and HIP~66198 (HD~118214), respectively. We used these standard stars to correct our science spectra for telluric absorption and for flux calibration. For the flux calibration we assumed that their continuum emission can be approximated by a blackbody with an effective temperature of $T_{\rm eff}=9600$~K for HIP~78017 and $T_{\rm eff}=9886$~K for HIP~66198 \citep{pastel}. Flats and arcs were taken after the science target.

We reduced the data using the Gemini/IRAF package with GNIRS specific tools \citep{gnirssoft}. The data reduction steps included preparation of calibration and science frames, processing and extraction of spectra, wavelength calibration, telluric correction, flux-calibration, and merging the different orders into a single, continuous spectrum. The spectral extraction width depends on the seeing during the observation, which on Mauna Kea is typically $0\farcs4 - 0\farcs8$ in the $K$ band, and was adjusted interactively for the telluric standard star and the science source and for each of the spectral orders to include all the flux in the spectral trace. A local averaged background flux was fitted and subtracted from the total source flux.

\subsubsection{IRTF~3~m} \label{irtf}

We observed Mrk~817 at the IRTF during two observing seasons covering the periods of 2021 Feb 12 to 2021 Sep 4 and 2022 Feb 3 to 2022 Jun 22. Of the 54 scheduled observing windows, we lost 13 due to weather and engineering issues. We used the SpeX spectrograph \citep{Ray03} in the short cross-dispersed mode (SXD, \mbox{$0.7-2.55~\mu$m}) and the $0\farcs3 \times 15\arcsec$ slit oriented at the parallactic angle. This set-up gives an average spectral resolution of $R=2000$ (FWHM $\sim 150$~km~s$^{-1}$). The on-source exposure time was usually $32\times120$~s taken with the usual ABBA pattern. The seeing on Mauna Kea, Hawaii, is typically $0\farcs4 - 0\farcs8$ in the $K$ band. 

With the same set-up, we observed the nearby A0~V star HD~121409, which we used to correct our science spectrum for telluric absorption and for flux calibration. Flats and arcs were taken mostly after the science target. We reduced the data using Spextool, which carries out all the procedures necessary to produce fully reduced spectra. This includes preparation of calibration frames, processing and extraction of spectra, wavelength calibration, telluric correction, flux-calibration, and merging the different orders into a single, continuous spectrum. As for the ARC spectra (see Section~\ref{arc}), we used the optimally weighted extraction method.

\subsubsection{Quasi-simultaneous near-IR spectra} \label{simspectra}

For 20 epochs, we have near-IR spectra obtained on the same day with two different telescopes: 12 \mbox{Gemini/ARC}, 4 Gemini/IRTF, 4 ARC/IRTF spectral pairs. Unfortunately, we do not have epochs obtained on the same day with all three telescopes. We used these spectral pairs to assess systematics. In particular, we used them to determine differences in the host galaxy flux contribution and to assess the reliability of the continuum spectral shapes in the $H$ and $K$ bands, which are important for the dust temperature blackbody fits.

\subsection{The near-IR photometry} \label{wircphotometry}

We obtained photometric data at the Palomar Observatory, California, with the 5~m Hale telescope from 2021 Jan 6 to 2022 May 31. Eleven partial nights were used for observations with the {\it Wide-field Infrared Camera} \citep[WIRC;][]{wirc}, which provides a plate scale of $0\farcs25$~pixel$^{-1}$. We used the $J$, $H-Cont$, and $K-Cont$ narrow-band filters, with the latter two continuum filters spanning the usual spectral-line-free regions of the typical $H$ and $K$ passbands\footnote{\url{https://sites.astro.caltech.edu/palomar/observer/200inchResources/WIRC/wirc_filters.html}}, and exposure times were tuned to avoid saturation of the AGN. The data were taken in a dithered mode with 15\arcsec offsets. The images were then subtracted in pairs to remove dark and sky contributions and then shifted and stacked for the final analysis images. The conditions were photometric on the nights of the observations but the seeing varied significantly and is captured in the FWHM of the PSF reported for each night in Table~\ref{wirc}.

The photometry for the AGN contribution was done with two different approaches. The preferred way is to do PSF fitting so that the central point source of the AGN would be isolated from the host galaxy. The PSF fitting used the nearby star 2MASS~J14361655+5847550, which lies $46\arcsec$ away, and so the 2MASS $J$, $H$, and $K_s$ photometric calibration is used \citep{2MASS}. The PSF fitting was achieved using the specialized IDL-package StarFinder \citep{Diolaiti00}, which is optimized for extracting PSFs from crowded fields and so is well suited to extract the AGN point source contribution from the Mrk~817 host galaxy. Nevertheless, the PSF fitting did not converge for all images. So we did a modified aperture photometry as well where we chose a photometry radius of 2 pixels and a sky subtraction ring of 2-4 pixels for both Mrk~817 and 2MASS~J14361655+5847550 so that the flues were dominated by the central region of both the PSF and AGN. Then by taking the ratio between the two fluxes achieved a more consistent photometric precision as evidenced by the same variability trends in all filters, but with increasing relative changes between each epoch of observation consistent with the decrease in the host galaxy contribution. The final photometric fluxes and their errors from the two methods are listed in Table~\ref{wirc}.

\subsection{The $z_s$-band photometry} \label{lcophotometry}

As part of the STORM~2 campaign, intensive ground-based imaging observations were obtained from several facilities covering the period of 2020 Nov - 2022 Feb in seven bands ($u$, $B$, $V$, $g$, $r$, $i$ and $z_s$). A large portion of these observations were obtained with the 1~m robotic telescope network of the Las Cumbres Observatory \citep[LCO;][]{LCOGT} and the Dan Zowada Memorial Observatory, New Mexico. LCO and Zowada were used to continue monitoring Mrk~817 as part of the Extended Campaign. Details of the observations, data reduction and inter-calibration between the observations from the different telescopes are given in \citet{Storm2-paper1} and in Montano et al. (in prep.). We used the light-curves obtained with the $g$ and $z_s$ filters, which have a wavelength width of 1500~\AA~and 1040~\AA~around their central wavelength of 4770~\AA~and 8700~\AA, respectively (Fig. \ref{zbandlcurves}, top two panels). The $z_s$ filter overlaps in wavelength with the IRTF and Gemini near-IR spectra, but the ARC near-IR spectra do not cover it sufficiently to permit comparisons (see Section~\ref{fluxscale}).

\subsection{The Integral Field Unit (IFU) near-IR spectroscopy} \label{ifu}

On 2022 Jun 4, we obtained an observation with the Near-Infrared Integral-Field Spectrograph \citep[NIFS;][]{nifs} on the Gemini North 8~m in queue mode (Program ID: \mbox{GN-2022A-Q-230}). We used the Z grating, which covers the wavelength range of $0.94 - 1.15$~$\mu$m with a resolving power $R = 4990$. For Mrk~817, this includes the strong \SIII~$\lambda 9531$ narrow emission line and the \HeI~1.08~$\mu$m broad emission line. The NIFS field-of-view (FOV) of $3\arcsec \times 3\arcsec$ is covered by spatial pixels (spaxels) with a size of $0\farcs103$ across slices and $0\farcs04$ along slices. Unfortunately, the Adaptive Optics (AO) system \mbox{ALTAIR} was unavailable and our obervation was taken in natural seeing with a FWHM$\sim 0\farcs45$.

We obtained 12 individual datacubes with an exposure time of 600~s each, which were reduced with the Python-based package Nifty4gemini. Telluric standard stars were observed before (HIP~67848) and after the science target (HIP~76376), which were used also for flux calibration. Lamp exposures were obtained for flat-fielding and wavelength calibration. For this paper, we constructed 2D maps only for the \SIII~$\lambda 9531$ emission line (Fig.~\ref{nifs}).

\subsection{The HST/STIS spectroscopy} \label{stis}

As part of the STORM~2 campaign, we obtained six low-resolution spectra with the Space Telescope Imaging Spectrograph \citep[STIS;][]{stis} on-board {\it HST} with a continuous wavelength coverage from the far-UV to the near-IR ($1600-10200$~\AA). The STIS visits were evenly distributed throughout the campaign (2020 Dec 3, 2020 Dec 18, 2021 Apr 18, 2021 Jul 25, 2022 Jan 2 and 2022 Feb 24) and used the $0\farcs2 \times 52\arcsec$~slit. For the longer wavelengths, we used the STIS CCD with the gratings G430L and G750L. Three 30~s exposures with each grating covered $2950–5700$~\AA~and $5300–10200$~\AA, respectively. More details are given in \citet{Storm2-paper1}, \citet{Storm2-paper2} and Plesha et al. (in prep.). The {\it HST} data products from AGN STORM~2 are available at MAST DOI 10.17909/N734-K698 \citep{Storm2-MAST}.

\subsection{The absolute spectral flux scale} \label{fluxscale}

\begin{figure}
\centerline{
\includegraphics[scale=0.9]{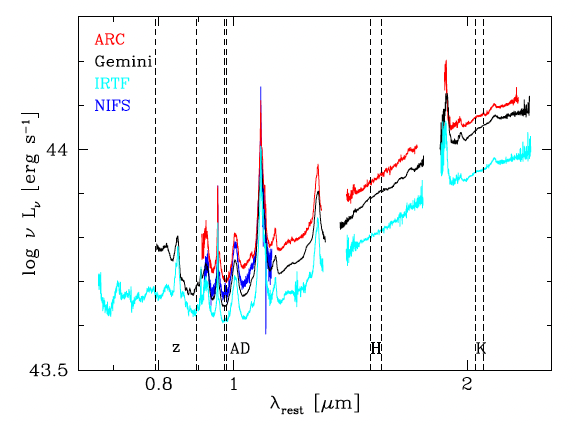}
}
\caption{\label{meanspectra} The mean spectra for the IRTF (cyan), Gemini North (black) and ARC data sets (red) derived after the application of the \prepspec~photometric correction factors and shown as luminosity versus rest-frame wavelength. The NIFS spectrum is also plotted (blue). The vertical dashed lines indicate the wavelength regions used to derive the spectral light-curves for the $z_s$ band (Fig.~\ref{zbandlcurves}), the accretion disk (Fig.~\ref{disklc}) and the dust $H$ and $K$~bands (Fig.~\ref{HKbandlc}).}
\end{figure}

\begin{figure}
\centerline{
\includegraphics[scale=0.85]{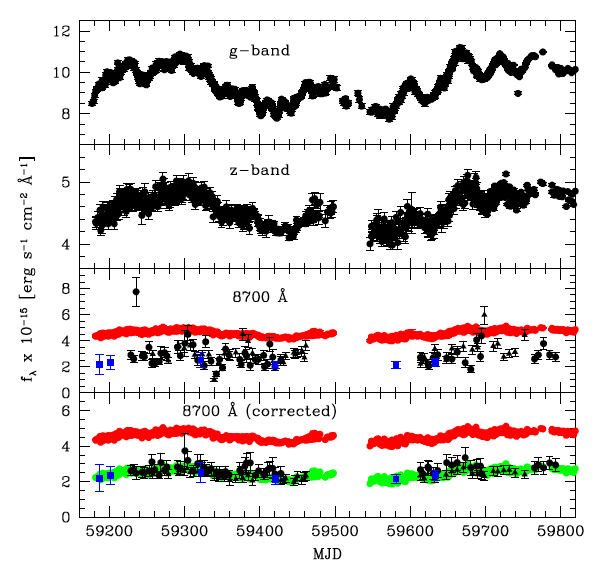}
}
\caption{\label{zbandlcurves} Top two panels: Photometric $g$ and $z_s$~band light-curves during the STORM~2 campaign from 2020 Nov 24 (MJD 59177) to 2022 Feb 24 (MJD 59634) and the beginning of the Extended Campaign. Bottom two panels: Gemini~North (black filled circles) and IRTF (black filled triangles) near-IR spectral light-curve around the observed wavelength of $8700$~\AA, both original and corrected using \prepspec~photometric correction factors based on the \SIII~$\lambda 9531$~narrow line. The original $z_s$ band light-curve (red filled circles) and the $z_s$ band light-curve corrected for a constant host galaxy contribution (green filled circles) to match the HST/STIS spectral fluxes (blue filled squares) are also shown.}
\end{figure}

In order to derive meaningful spectral light-curves, we must achieve an accurate absolute flux calibration of the spectra. As \citet{vanGron92} showed, matching the profile rather than the flux of a strong and non-variable emission line can yield a photometric alignment of spectra at the $\sim 1–5$ per cent level. Similar to the approach chosen by optical spectroscopic reverberation mapping campaigns, we based our absolute spectral flux scale on a strong narrow emission line from a forbidden transition. Such emission lines are expected to remain constant during the campaign, since they are produced in gas that is located at large distances from the ionizing source and that has an electron density low enough for recombination timescales to be large. Following \citet{L19, L23a}, we chose the \SIII~$\lambda 9531$~line since it is the strongest narrow forbidden emission line in the near-IR. It is blended with the Pa$\epsilon$~broad emission line, but in Mrk~817 the two can be easily separated (Fig.~\ref{irspec}). 

We used the \prepspec~routine \citep{Horne20} to determine the photometric correction factors for our near-IR spectroscopy. In short, \prepspec~models all emission lines and the total continuum and matches the profiles of selected narrow emission lines to derive the time-dependent flux scaling factor. It assumes that the majority of the spectra are photometric and so holds the median photometric correction factor at one. Since the three spectral data sets have different spectral resolutions, we first intercalibrated them separately and then tied them together assuming that their {\it median} \SIII~$\lambda 9531$~line fluxes are equal. In practice, this meant dividing the resulting \prepspec~photometric correction factors for the IRTF spectra by a factor of 1.32 to align their median \SIII~line fluxes with those of the other two data sets. The final spectral scaling factors and their errors are listed in Table \ref{lightcurves}. Although we find that in a few cases relatively large photometric correction factors of $\sim 2-8$ are required, their distribution has a mean and dispersion of $1.01 \pm 0.67$ and a median of 0.93. For the three data sets separately, the means, dispersions and medians are $0.80 \pm 0.28$ and 0.76 (IRTF), $1.00 \pm 0.23$ and 1.00 (\mbox{Gemini}), and $1.15 \pm 1.05$ and 1.00 (ARC). Fig.~\ref{meanspectra} shows the resulting mean IRTF, Gemini and ARC spectra, which have \SIII~$\lambda 9531$~line fluxes of $(1.44\pm0.07) \times 10^{-14}$, $(1.39\pm0.10) \times 10^{-14}$ and $(1.72\pm0.05) \times 10^{-14}$~erg~s$^{-1}$~cm$^{-2}$, respectively.

Since \prepspec~is agnostic to the true flux scale of the median, we tested our results with the NIFS IFU observation. Since its PSF FWHM is similar to the slit width used for the Gemini~North spectra, we compare the NIFS nuclear spectrum to the mean Gemini spectrum in Fig.~\ref{meanspectra}. The two match well in flux and shape, although the NIFS observation samples only a single epoch of the variable continuum flux. We also derived the \SIII~$\lambda 9531$~line flux of the unresolved, nuclear component from the NIFS observation, which was \mbox{$(1.57\pm0.11) \times 10^{-14}$~erg~s$^{-1}$~cm$^{-2}$}, similar to the mean values for the near-IR spectra. However, we found that the \SIII~$\lambda 9531$~emission is extended in Mrk~817 and discuss its morphology further in Section~\ref{extdust}. The curve of growth shows that the integrated flux of the extended component reaches its maximum at a radius of $\sim 0\farcs8$ from the center and constitutes $\sim 20\%$ of the total flux (Fig.~\ref{nifsradius}). Integrating the \SIII~$\lambda 9531$~line flux over the entire NIFS FOV, we measure \mbox{$(1.80\pm0.12) \times 10^{-14}$~erg~s$^{-1}$~cm$^{-2}$}. This extended flux is expected to be fully enclosed in the ARC spectra, which explains the slightly larger \SIII~line flux of the mean ARC spectrum compared to the mean IRTF and Gemini spectra.  

We further used the HST/STIS spectroscopy in combination with the photometric $z_s$ band light-curve to test the accuracy of the \prepspec~results. In Fig.~\ref{zbandlcurves} (bottom two panels), we show the IRTF and Gemini fluxes corresponding to the $z_s$ band before and after the correction. Since the wavelength range of the ARC spectra does not sufficiently cover the $z_s$ band, they are not included. If we subtract from the $z_s$~band photometric light-curve a constant host galaxy contribution of $\sim 50\%$ of the total flux (green circles in bottom panel), it matches all STIS spectral fluxes. This light-curve then matches also the ground-based spectroscopy, which means that any difference in host galaxy contribution to the HST, IRTF and Gemini spectra must be negligible. We present a further test of the reliability of our absolute flux calibration at the end of Section~\ref{lcurves}.

The errors on the \prepspec~correction factors, which represent the nominal final errors on our absolute flux calibration, are $\sim 5\%$ on average (Table~\ref{rmstable}). This is in line with the accuracy expected for ground-based spectroscopy when matching line profile shapes instead of line fluxes, particularly for high $S/N$ spectra such as ours \citep[see the discussion in][]{Faus17}. To assess if the errors on the \prepspec~scaling factors could be underestimated, we also used the publicly available \mapspec~routine \citep{Faus17}\footnote{\url{https://github.com/mmfausnaugh/mapspec}}. \mapspec~matches the profile of a single emission line and does not provide errors on its scaling factors. After accounting for the fact that \prepspec~keeps the median of the scaling factors at one, the resulting \mapspec~factors were on average within the $1\sigma$ errors of the \prepspec~factors.

\subsection{The spectral light-curves} \label{lcurves}

\begin{figure}
\centerline{
\includegraphics[scale=0.85]{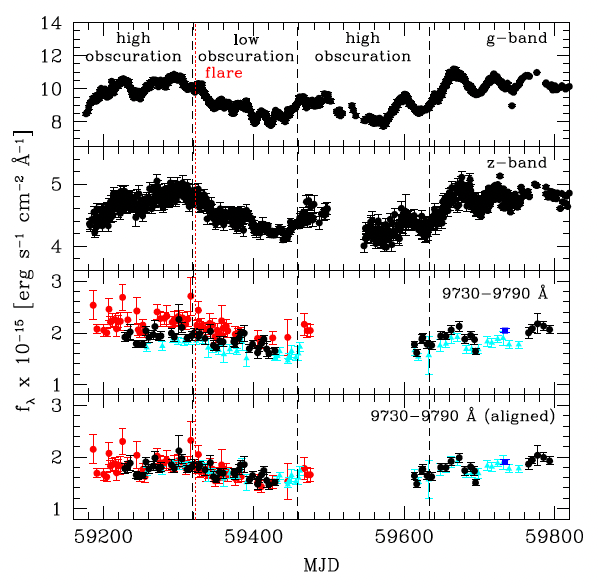}
}
\caption{\label{disklc} Top two panels: Photometric $g$ and $z_s$~band light-curves during the STORM~2 campaign (2020 Nov 24 to 2022 May 1) and the Extended Campaign. Bottom two panels: Near-IR spectral light-curve in the rest-frame wavelength region of $\lambda=9730-9790$~\AA, which is dominated by accretion disk contributions, for the Gemini~North (black filled circles), IRTF (cyan filled triangles) and ARC spectra (red filled circles). The NIFS spectrum is also plotted (blue filled square). We attribute the offset to higher fluxes of the ARC spectra relative to the Gemini North spectra and of these relative to the IRTF spectra to differences in the host galaxy light contribution. Subtracting this constant component gives a well-aligned near-IR spectral light-curve (bottom panel). The vertical black dashed lines mark the periods of high- and low-obscuration until the end of the STORM~2 campaign, as defined by \citet{Storm2-paper3}. The vertical red dotted line marks the date of the X-ray flare (MJD=59329).}
\end{figure}

\begin{figure}
\centerline{
\includegraphics[scale=0.85]{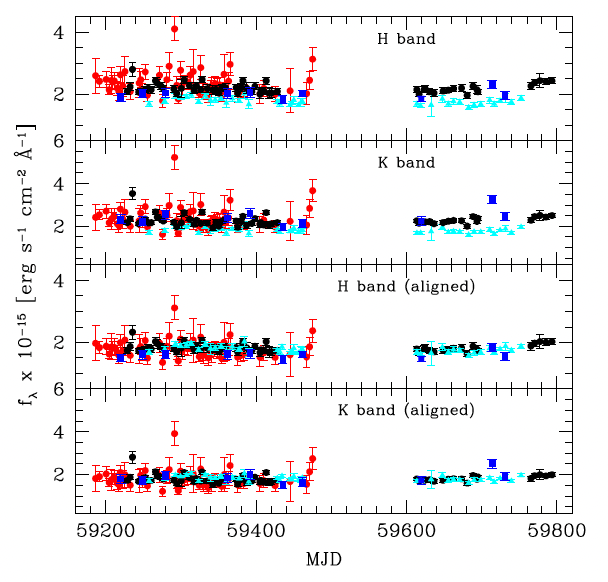}
}
\caption{\label{HKbandlc} Top two panels: Near-IR spectral light-curves for the rest-frame wavelength regions of $\lambda=1.50-1.55~\mu$m (H~band) and $\lambda=2.05-2.10~\mu$m (K~band) for the \mbox{Gemini}~North (black filled circles), IRTF (cyan filled triangles) and ARC spectra (red filled circles). The WIRC photometry is also plotted (blue filled squares). We attribute the offset of the IRTF spectra to lower fluxes to an additional dust component contributing to the Gemini~North and ARC spectra. Subtracting this constant dust component gives well-aligned near-IR spectral light-curves (bottom two panels).}
\end{figure}

We derived light-curves for the UV/optical AGN component assumed to be the accretion disk and for the thermal component attributed to hot dust from the near-IR spectra. Ideally, the accretion disk flux should be measured at the shortest wavelengths, where the contribution from the hot dust component is minimal. However, the $S/N$~ratio of the near-IR spectra decreases towards the shortest wavelengths and line blending for $7200 \lesssim \lambda \lesssim 9750$~\AA~significantly hampers continuum-level determination. Therefore, we measured the accretion disk flux in the 60~\AA~wide rest-frame wavelength region of $\lambda=9730-9790$~\AA, which lies between the two broad Pa$\epsilon$ and Pa$\delta$ hydrogen emission lines (Fig.~\ref{irspec}, but see also Fig.~\ref{meanspectra}) and in the observer's frame lies close to the typical wavelength of the photometric $Y$~band. Table~\ref{lightcurves} lists the mean flux and root-mean-square (rms) and the light-curve is displayed in Fig.~\ref{disklc} in comparison to the $g$ and $z_s$ band accretion disk light-curves. The average rms values are $\sim 1-3\%$  (Table~\ref{rmstable}), which means that the error budget on the light-curve is dominated by the errors in the \prepspec~scaling factors.

It is immediately evident from Fig.~\ref{disklc} that the spectra from the different telescopes are offset to larger fluxes for larger slit sizes (i.e. the ARC spectra are offset to larger fluxes relative to the Gemini spectra which in turn are offset to larger fluxes relative to the IRTF spectra). This offset, which is also apparent in Fig.~\ref{meanspectra} for the mean spectra, is most likely due to an increased host galaxy flux contribution within the larger slits. The HST/WFC image analysis of \citet{Bentz09} showed that the host galaxy of Mrk~817 has a relatively compact bulge with an effective radius of $\sim 0\farcs5$ and a radial extent of $\sim 3\arcsec$. In the $V$ band, the luminosity of the bulge constitutes $\sim 5\%$ of the total galaxy light, with the galaxy disk component dominating the flux at radii $\ga 0\farcs7$. Based on the simultaneous spectra (Section~\ref{simspectra}) and the mean spectra, we determined the average excess host galaxy light, which we then subtracted from the ARC and Gemini data in order to align them with the IRTF spectral light-curve (Fig.~\ref{disklc}, bottom panel). The excess flux was on average $\sim 7\%$ and $\sim 18\%$ of the total flux for the Gemini and ARC spectra, respectively. The rest-frame wavelength region of $\lambda=9730-9790$~\AA~is covered also by the NIFS spectrum and we added this data point to the light-curve. As Fig.~\ref{disklc} (bottom panel) shows, subtracting a similar host galaxy contribution as for the Gemini spectra aligns it with the light-curve.   

We derived the hot dust light-curve in two line-free, 500~\AA~wide rest-frame wavelength regions in the middle of the $H$ and $K$~bands spanning $\lambda=1.50-1.55~\mu$m and $\lambda=2.05-2.10~\mu$m, respectively (Fig.~\ref{meanspectra}). The values are listed in Table \ref{lightcurves} and the light-curves are displayed in Fig.~\ref{HKbandlc} (top two panels). It is immediately evident from Fig.~\ref{HKbandlc} that the IRTF spectra are offset to lower fluxes relative to both the Gemini and ARC spectra. Since the shifts are similar for both data sets and both bands, we can exclude pure stellar host galaxy light as the main cause. For example, the $H$ and $K$ band fluxes of the Sb galaxy template of \cite{Pol07} are different and lower by a factor of $\sim 2$ and $\sim 4$, respectively, than the host galaxy flux in the rest-frame $\lambda=9730-9790$~\AA~spectral range. Therefore, the observed offset for the IRTF spectra indicates that both the Gemini and ARC spectra include an additional dust component, which, based on the slit sizes, must be located on scales $\ga 0\farcs225$ or $\ga 140$~pc and so must be constant. Using the simultaneous spectra and the mean spectra, we obtain an average excess dust flux for the Gemini and ARC spectra of $\sim 25\%$ and $\sim 30\%$ of the total $H$ and $K$ band fluxes, respectively. Subtracting this constant component from the Gemini and ARC data aligns them with the IRTF spectral light-curves (Fig.~\ref{HKbandlc}, bottom two panels). The existence of an extended dust component is supported by the WIRC near-IR photometry, which we added to the spectral light-curves in Fig.~\ref{HKbandlc}. 

The extended, constant dust lies most likely within the compact bulge of the host galaxy. The deep $H$ band images of \citet{Bentz18} obtained with the WIYN High-Resolution Infrared Camera (WHIRC), which has a plate scale of $0\farcs0986$~pixel$^{-1}$, showed the bulge of Mrk~817 to be unusually bright in the near-IR. Whereas the bulge luminosity constitutes only $\sim 5\%$ of the total optical galaxy flux, the fraction increases to $\sim 20\%$ in the $H$ band. This is also evidenced by its unusual colors. While the color of the galaxy disk component of \mbox{$V - H = 2.76$~mag} is typical, the bulge had the reddest color in the sample of \citet{Bentz18} with \mbox{$V - H = 4.38$~mag}. Their estimated total $H$ band bulge flux corresponds to a luminosity of $L_{\rm bulge, H} \sim 2.3 \times 10^{43}$~erg~s$^{-1}$, which can account for the average flux offset we observe between the Gemini and ARC spectra and the IRTF data (Fig.~\ref{meanspectra}). We discuss this extended hot dust component further in Section~\ref{extdust}.

The tests in Section~\ref{fluxscale} showed that \prepspec~performs well in principle to correct our fluxes to an absolute scale, but there are no strong forbidden narrow emission lines in the $H$ and $K$ bands (see Fig.~\ref{irspec}) to verify that the telluric correction accounted well for the wavelength-dependent seeing effects across the entire wavelength range. Since the \prepspec~correction is heavily weighted towards the strong \SIII~$\lambda 9531$ line and the $9730-9790$~\AA~($Y$ band) continuum lies very close to it, the variability of this spectral light-curve should be representative of the true variability of the accretion disk component. We calculated its intrinsic variability relative to the mean flux, $F_{\rm var}$, following \citet{fvar} (see their eqs. 3 and 4). We find similar values for the three data sets of $\sim 4-6\%$ (Table~\ref{rmstable}), which are comparable to the intrinsic variabilities of the space-based {\it Swift} $V$ band \citep[$F_{\rm var} = 0.071$;][]{Storm2-paper4} and ground-based $z_s$ band light-curves ($F_{\rm var} = 0.051$). Since the calculation of $F_{\rm var}$ takes the errors into account, this result also shows that the error budget on the $Y$ band light-curve is reasonable. But the level of intrinsic variability is low and comparable to that of the combined uncertainties from the data and the \prepspec~correction.

Next, we calculated the $F_{\rm var}$ values for the $H$ and $K$ band light-curves as well as for the flux ratio between the $H$ and $K$ band fluxes, which is independent of the \prepspec~correction factors. No intrinsic variability was present for the IRTF $H$ band light-curve, whereas that for the Gemini North data was at the $\sim 4\%$ level and so comparable to the $Y$ band intrinsic variability. The ARC data set showed a large value of $\sim 15\%$, indicative of the errors being underestimated. Using only the simultaneous spectra (Section~\ref{simspectra}), we found that increasing the errors by a factor of $\sim 2$ brought the ARC $H$ band intrinsic variability close to that of the Gemini spectra. The intrinsic variability was higher for the $K$ band for all three data sets, but, again, the ARC data showed a relatively large $F_{\rm var}$ value. An increase of their errors by a factor of $\sim 2.5$ seemed to be necessary to reduce their intrinsic variability to that of the Gemini data. On the other hand, the three data sets had similar intrinsic variabilities in their $H/K$ band flux ratios of $\sim 2-6\%$, comparable to that of the $Y$ band light-curve.

\begin{deluxetable*}{lccccccccc}
\tablecolumns{10}
\decimalcolnumbers
\tablecaption{\label{rmstable} 
Properties of the near-IR spectral light-curves}
\tablehead{
\colhead{Telescope} & \colhead{cadence} & \colhead{PrepSpec} & \colhead{$Y$} & \colhead{$H$} & \colhead{$K$} & \multicolumn{4}{c}{$F_{\rm var}$} \\
& \colhead{(days)} & \colhead{(\%)} & \colhead{(\%)} & \colhead{(\%)} & \colhead{(\%)} & \colhead{$Y$} & \colhead{$H$} & \colhead{$K$} & \colhead{$H/K$}
}
\startdata
ARC          & 7 & 6 & 3 & 2 & 1 & 0.042 & 0.145 & 0.214 & 0.057 \\
Gemini North & 5 & 5 & 1 & 2 & 1 & 0.058 & 0.042 & 0.093 & 0.041 \\
IRTF         & 6 & 5 & 2 & 1 & 1 & 0.038 & $-$   & 0.015 & 0.021 \\
\enddata

\tablecomments{The columns are: (1) Telescope; (2) average cadence of the time-series; (3) mean $1\sigma$ percentage error on the \prepspec~correction factors; mean $1\sigma$ percentage error on the (4) $9730-9790$~\AA~($Y$ band), (5) $1.50-1.55~\mu$m ($H$ band), and (6) $2.05-2.10~\mu$m ($K$ band) continuum fluxes; intrinsic variability relative to the mean for the (7) $9730-9790$~\AA~($Y$ band), (8) $1.50-1.55~\mu$m ($H$ band), (9) $2.05-2.10~\mu$m ($K$ band) and (10) $H/K$ band flux ratio.}

\end{deluxetable*}

\section{The variability of the hot dust}

In order to study the variability of the hot dust, we need to separate its emission from that of the other variable components assumed to be present in the near-IR spectra of AGN, namely, the outer regions of the accretion disk and the diffuse emission from the BLR \citep{Kor01}. The advantage of our cross-dispersed near-IR spectroscopy over photometry is that it includes a large part of the near-IR SED and so simultaneously constrains these components. In Section~\ref{accdisclag}, we first investigate the variability of the continuum we attribute to the accretion disk and perform a spectral decomposition to isolate the hot dust spectrum in Section~\ref{dustspectrum}. The spectral decomposition has two main aims for the purpose of the dust reverberation mapping of Section~\ref{revsignal}: (i) to derive the light-curve of the hot dust and (ii) to derive the variations in the dust temperature.

\subsection{The accretion disk lag} \label{accdisclag}

\begin{figure}
\centerline{
\includegraphics[scale=0.3]{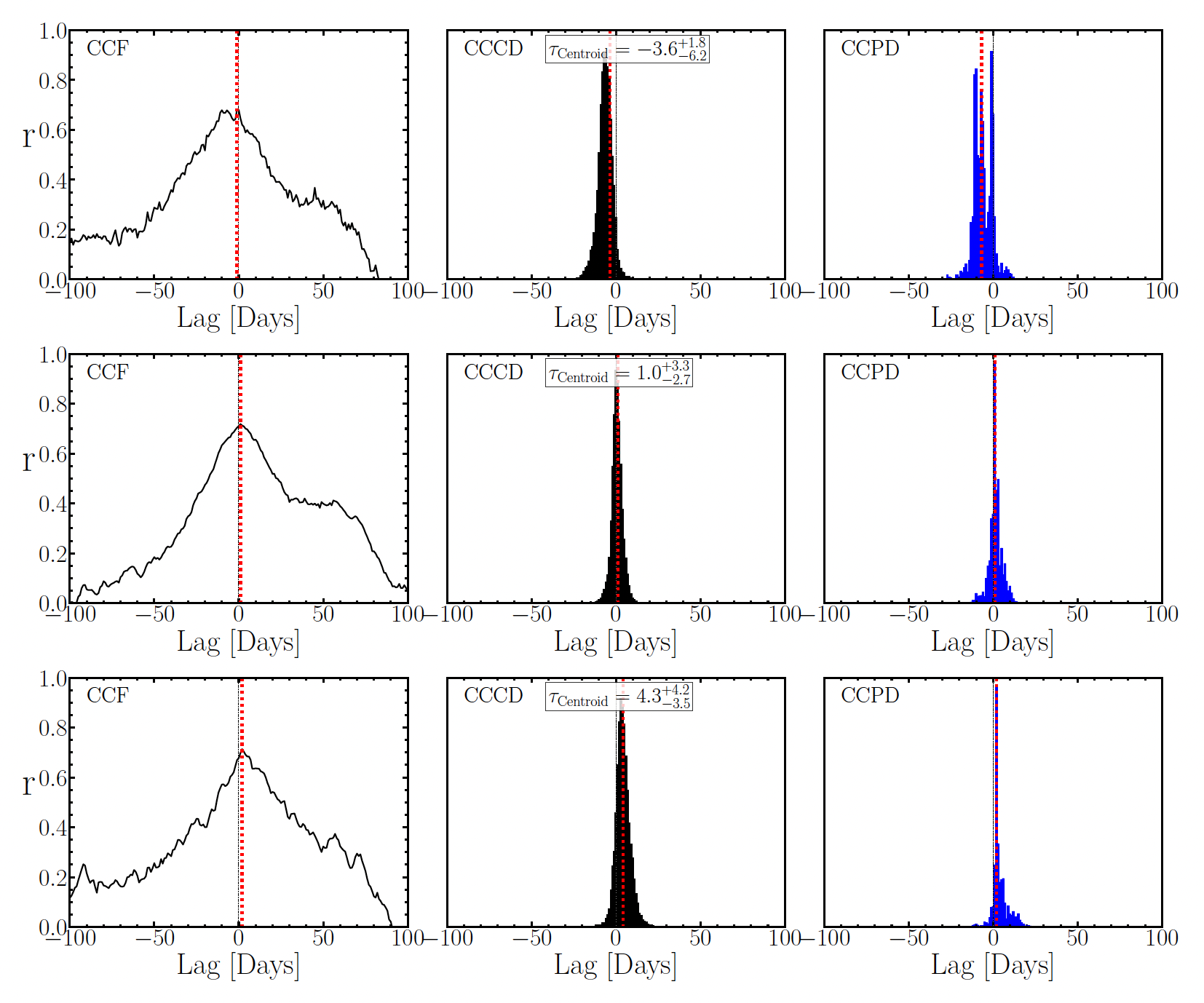}
}
\caption{\label{iccflagyband} 
Results of the PyCCF lag analysis for the rest-frame $\lambda=9730-9790$~\AA~light-curve, corresponding to the accretion disk. From top to bottom it was run versus the $z_s$-band, $g$-band and {\it Swift} UVW2 photometric light-curves. For each light-curve pair, we compute the distributions for the cross-correlation coefficient $r$, cross-correlation centroid (CCCD) and cross-correlation peak (CCPD). The lag measurements (vertical red dotted lines) and uncertainties are displayed at the top of the respective middle panels. For comparison, the locations of zero lag are also indicated (vertical black dotted lines).}
\end{figure}

Based on the near-IR radius-luminosity relationship and quasi-simultaneous optical and near-IR spectroscopy of high-luminosity type 1 AGN, the accretion disk emission is expected to dominate the continuum flux at wavelengths $\la 1~\mu$m \citep{L11b, L11a}. Therefore, our near-IR spectra allow us to sample this component up to the largest wavelengths and thus extend the wavelength range of the UV/optical continuum lags of the STORM~2 campaign. 

We performed a time-series analysis of the accretion disk light-curve shown in Fig.~\ref{disklc} versus the $z_s$-band, $g$-band and {\it Swift} UVW2 photometric light-curves using the Interpolated Cross-correlation Function \citep[ICCF;][]{Gas87, Pet04} as implemented in the publicly available PyCCF code \citep{ICCF}. PyCCF calculates the Pearson coefficient $r$ between two light-curves shifted by a range of time lags $\tau$ using linear interpolation to match the shifted light-curves in time. We obtained centroid lags of $\tau = -4^{+2}_{-6}$~days, $1\pm3$~days and $4\pm4$~days for the $z_s$, $g$ and {\it Swift} UVW2 bands, respectively (Fig.~\ref{iccflagyband}). The centroid of the ICCF is computed using points surrounding the maximum of the correlation coefficient $r_{\rm max} (r > 0.8~r_{\rm max})$, and the lag uncertainty is computed from Monte Carlo (MC) iterations of flux sampling and random subset sampling \citep[FS/RSS method;][]{Pet98b}. To estimate the uncertainties, we used 20,000 MC iterations of FS/RSS for PyCCF to generate a cross-correlation centroid distribution (CCCD) and a cross-correlation peak distribution (CCPD), which gives the distribution of measured lags in all of the MC realizations. The upper and lower lag uncertainties are measured from the 16th and 84th percentiles of the primary peak. 

The $9730-9790$~\AA~lag versus the {\it Swift} UVW2 band is consistent with expectations based on the normalizations for the UV/optical lag relationships found by \citet{Storm2-paper7} (see Section~\ref{target}). The negative lag found versus the $z_s$ band could indicate that this photometric filter has significant contribution from variable emission at larger scales. As Fig.~\ref{irspec} shows, the contaminants might be from the BLR, as often found for the $u$-band lag excess in AGN \citep{Edelson19}, since within its rest-frame wavelength range for Mrk~817 (of $\lambda=7934-8943$~\AA) the higher-order Paschen lines (Pa11 and higher) form a pseudo-continuum and strong \OI~emission and part of the Paschen jump expected from the BLR DC emission are also present.

\subsection{The hot dust spectrum} \label{dustspectrum}

\begin{figure}
\centerline{
\includegraphics[scale=0.85]{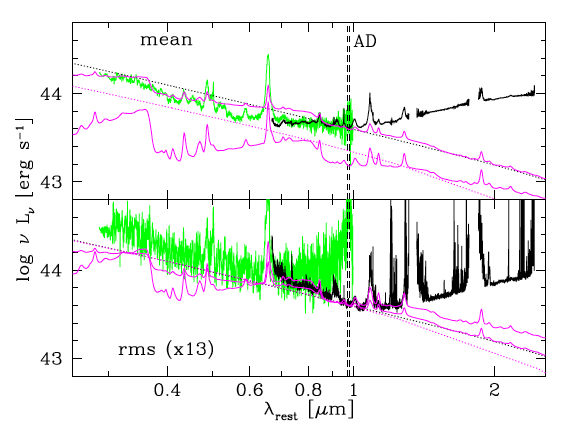}
}
\caption{\label{rmsspecdc} Top panel: Mean IRTF (black) and STIS spectrum (green). The vertical dashed lines indicate the rest-frame wavelength region of $\lambda=9730-9790$~\AA~used to derive the accretion disk spectral light-curves of Fig.~\ref{disklc}. The spectrum of an accretion disk with an outer radius of $r_{\rm out}=10^4 r_{\rm g}$ that reproduces the flux in this spectral region significantly overpredicts the UV/optical continuum flux (black dotted curve). Assuming instead the model of \citet{Storm2-paper10} (thick magenta solid line), which is the sum of BLR DC emission (thin magenta solid curve) and a fainter accretion disk with an outer radius of $r_{\rm out}=2160 r_{\rm g}$ (magenta dotted curve), can qualitatively reproduce the mean spectrum at UV/optical wavelengths. We note that the continuum spectrum of the \citet{Storm2-paper10} model and the accretion disk with an outer radius of $r_{\rm out}=10^4 r_{\rm g}$ are identical at near-IR wavelengths. Bottom panel: As in top panel for the variable (rms) spectrum scaled by a factor of 13 to match the $9730-9790$~\AA~flux of the mean. The models were normalised in this spectral region to compare their shapes. The accretion disk spectrum approximates best the variable UV/optical continuum flux.} 
\end{figure}

\begin{figure}
\centerline{
\includegraphics[scale=0.85]{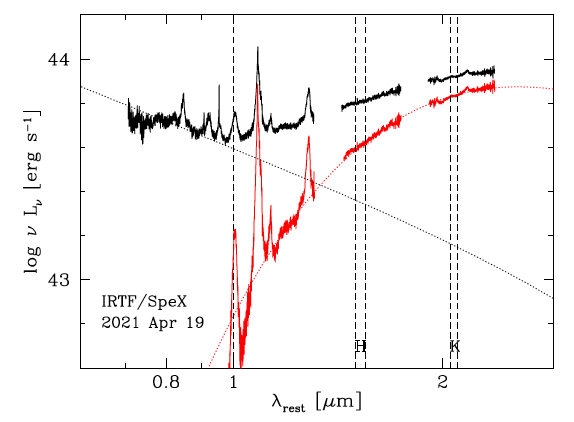}
}
\caption{\label{irsed} IRTF/SpeX near-IR spectrum from 2021 Apr 19 shown as luminosity versus rest-frame wavelength (black spectrum). We decomposed the continuum into two components, namely, an accretion disk spectrum that approximates the wavelength range $\la 1~\mu$m (black dotted curve) and still dominates at $1~\mu$m (vertical dashed line) and hot dust emission (red spectrum). We fitted the hot dust continuum with a blackbody spectrum, resulting in this example in a best-fit temperature of $T=1430$~K (red dotted curve). The wavelength regions used to derive the $H$ and $K$~band spectral light-curves are marked by vertical dashed lines.
}
\end{figure}

\begin{figure}
\centerline{
\includegraphics[scale=0.85]{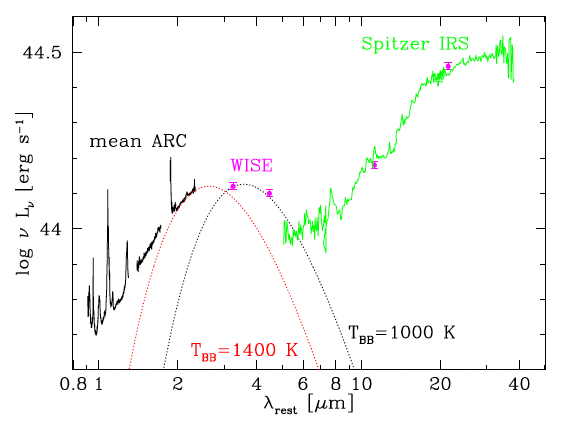}
}
\caption{\label{spitzer} Non-simultaneous near- to mid-IR spectral energy distribution for Mrk 817. The mean ARC/TripleSpec near-IR spectrum (black solid curve) and WISE W1 and W2 data (magenta filled circles) indicate that the hottest dust has a range of blackbody temperatures spanning $T \sim 1400$~K (red dotted curve) to $T \sim 1000$~K (black dotted curve). The {\it Spitzer} IRS spectrum (green solid curve) and WISE W3 and W4 data points (magenta filled circles) sample a much cooler ($T \sim 100-300$~K) dust component.
}
\end{figure}

Based on polarization studies \citep{Kish08} and observations of hot-dust-poor type 1 AGN \citep{Hao10}, the accretion disk emission spectrum is expected to extend to at least $\sim 2$~$\mu$m. The spectral slope of the BLR DC emission is generally much flatter than that of the accretion disk, and it is predicted to display strong emission jumps at the edges of the hydrogen series \citep[e.g.][]{Lawther18, Kor19}. In the near-IR wavelength range covered by our spectra, we expect to detect mainly the Paschen jump at a rest-frame wavelength of $\sim 8206$~\AA, since the Brackett jump at $\sim 1.46$~$\mu$m will be overwhelmed by emission from the hot dust. The DC emission is estimated to contribute $\sim 10-40\%$ to the total UV/optical continuum flux, with its flux directly dependent on the strength of the BLR, but beyond $\sim 1$~$\mu$m, where it is attributed to free-free continuum emission, it can dominate over the accretion disk \citep{Netzer20}. The spectral slope of the total spectrum can then mimick that of an accretion disk, in part because the electron scattering component produces a copy of the incident ionising continuum, which is particularly strong in the UV.

In order to assess which variable AGN components contaminate the $H$ and $K$ band fluxes of the hot dust, we created for the HST/STIS and IRTF data that together cover a very large wavelength range from the UV/optical to the near-IR mean and variable (rms) spectra following \citet{Pet04}. While the calculated rms spectrum includes the noise, it is dominated by the intrinsic variations given the relatively small errors. We find that a standard optically thick, geometrically thin accretion disk \citep{Shak73} for the black hole mass from Section \ref{target} and with an outer radius large enough to have maximum emission in the $K$~band (of $r_{\rm out}=10^4 r_{\rm g} \sim 20$~light-days) significantly overpredicts the mean UV/optical continuum flux, if matched to the $9730-9790$~\AA~flux (Fig.~\ref{rmsspecdc}, top panel). However, this accretion disk spectrum reproduces well the spectral shape of the variations in the UV/optical/near-IR regime up to wavelengths slightly beyond $1~\mu$m (Fig.~\ref{rmsspecdc}, bottom panel). A similar result was found also by \citet{Storm2-paper4} for the rms spectrum isolated from a flux-flux analysis of the UV/optical light-curves (see their Fig.~11). Assuming instead the model of \citet{Storm2-paper10}, which is the sum of BLR DC emission and a fainter and smaller accretion disk with an outer radius of $r_{\rm out}=2160 r_{\rm g} \sim 4$~light-days, we can qualitatively reproduce the mean spectrum at UV/optical wavelengths, however, the variable spectral flux remains mostly underpredicted. Therefore, since the relevant aspect for the isolation of the hot dust spectrum for the purpose of reverberation mapping is the continuation of the spectrum into the near-IR for the component dominating the {\it variable} flux, we assumed only an accretion disk. However, we note that the continuum spectrum of the \citet{Storm2-paper10} model and that of the accretion disk with an outer radius of $r_{\rm out}=10^4 r_{\rm g}$ are identical at near-IR wavelengths $\ga 1~\mu$m. 

Following \citet{L19}, we first approximated the rest-frame wavelength range $\la 1~\mu$m with the accretion disk spectrum, which we then subtracted from the total spectrum. Thus the accretion rates were obtained directly from a scaling of the model to the individual spectra. We then fit the resulting hot dust spectrum at wavelengths $>1~\mu$m excluding emission lines with a single blackbody, representing emission by large dust grains. \citet{L23b} found for a sample of $\sim 40$ type 1 AGN that luminosity-based dust radii assuming a blackbody agreed within a factor of $\sim 2$ with dust radii obtained by independent methods such as reverberation mapping and interferometry, whereas estimates assuming a modified blackbody representing small-grain carbon or silicate dust were much larger (by factors of $\sim 6-10$). Large dust grains (with sizes of a few $\mu$m) appear to dominate the hot dust composition in AGN and, therefore, we did not consider a modified blackbody. In Fig.~\ref{irsed}, we show an example, and Table \ref{lightcurves} lists the $H$ and $K$ band dust fluxes and blackbody dust temperatures resulting from the spectral decomposition. 

The IRTF and Gemini spectra have a sufficiently large wavelength coverage in the blue to allow for a reliable anchoring of the accretion disk spectrum and they yield similar mean dust temperatures of $\langle T \rangle = 1396 \pm 28$ and $1393 \pm 32$~K, respectively. Here and in the remainder of the paper we give the mean and the dispersion around the mean. As we have shown in Section~\ref{lcurves}, the ARC spectra have the largest host galaxy flux contribution. Although the host galaxy emission peaks at a rest-frame wavelength of $\sim 1~\mu$m and then steeply decreases towards longer wavelengths \citep[see, e.g., Fig. 6 in][]{L11a} and so is not expected to affect the best-fit dust temperatures, we can test this assumption. A significant contribution from the host galaxy would manifest itself in apparently higher dust temperatures due to excess flux mainly at the shortest wavelengths. For the ARC spectra, we get a mean blackbody dust temperature of $\langle T \rangle = 1394 \pm 30$~K, which is consistent with our results for the IRTF and Gemini spectra. 

Although the near-IR spectra span a large wavelength range and have a high $S/N$, which lead to small formal errors on the dust temperatures, we used the quasi-simultaneous spectral pairs to assess the systematic uncertainties in the spectral shape. We found that their dust temperatures are consistent with each other within $2\sigma$, with only 4/20 pairs showing a $3\sigma$ difference of $\Delta T \sim 40 - 60$~K. The best-fits gave reduced $\chi^2$ values close to one, indicating that a second blackbody is not required for the wavelength range of our near-IR spectra. However, the broader near- to mid-IR SED of Mrk~817 (Fig.~\ref{spitzer}), shows that the hot dust does span a considerable temperature range of $T \sim 1000 - 1400$~K and much cooler dust (of $T \sim 100 - 300$~K) is also seen at the still longer ($\ga 10~\mu$m) wavelengths sampled by the {\it WISE} and {\it Spitzer} space telescopes.

\subsection{The dust reverberation signal} \label{revsignal}

\begin{figure}
\centerline{
\includegraphics[scale=0.85]{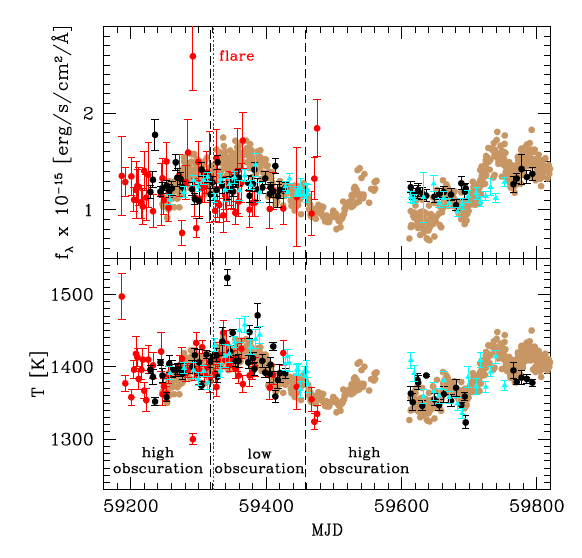}
}
\caption{\label{templc} 
The temporal variability of the hot dust flux in the $H$ band (top panel) and the dust temperature (bottom panel) resulting from the spectral decomposition. We plot 1$\sigma$ errors. Symbols are as in Fig.~\ref{disklc}. The dust temperature light-curve follows that of the photometric $z_s$~band light-curve shifted in time by the ICCF lag $\tau = 65$~days (brown curve; see also Fig.~\ref{iccflagtemp}), whereas the dust flux light-curve does not have a clear response to it. The vertical black dashed lines mark the periods of high- and low-obscuration until the end of the STORM~2 campaign. The vertical red dotted line marks the date of the X-ray flare (MJD=59329).}
\end{figure}

\begin{figure}
\centerline{
\includegraphics[scale=0.3]{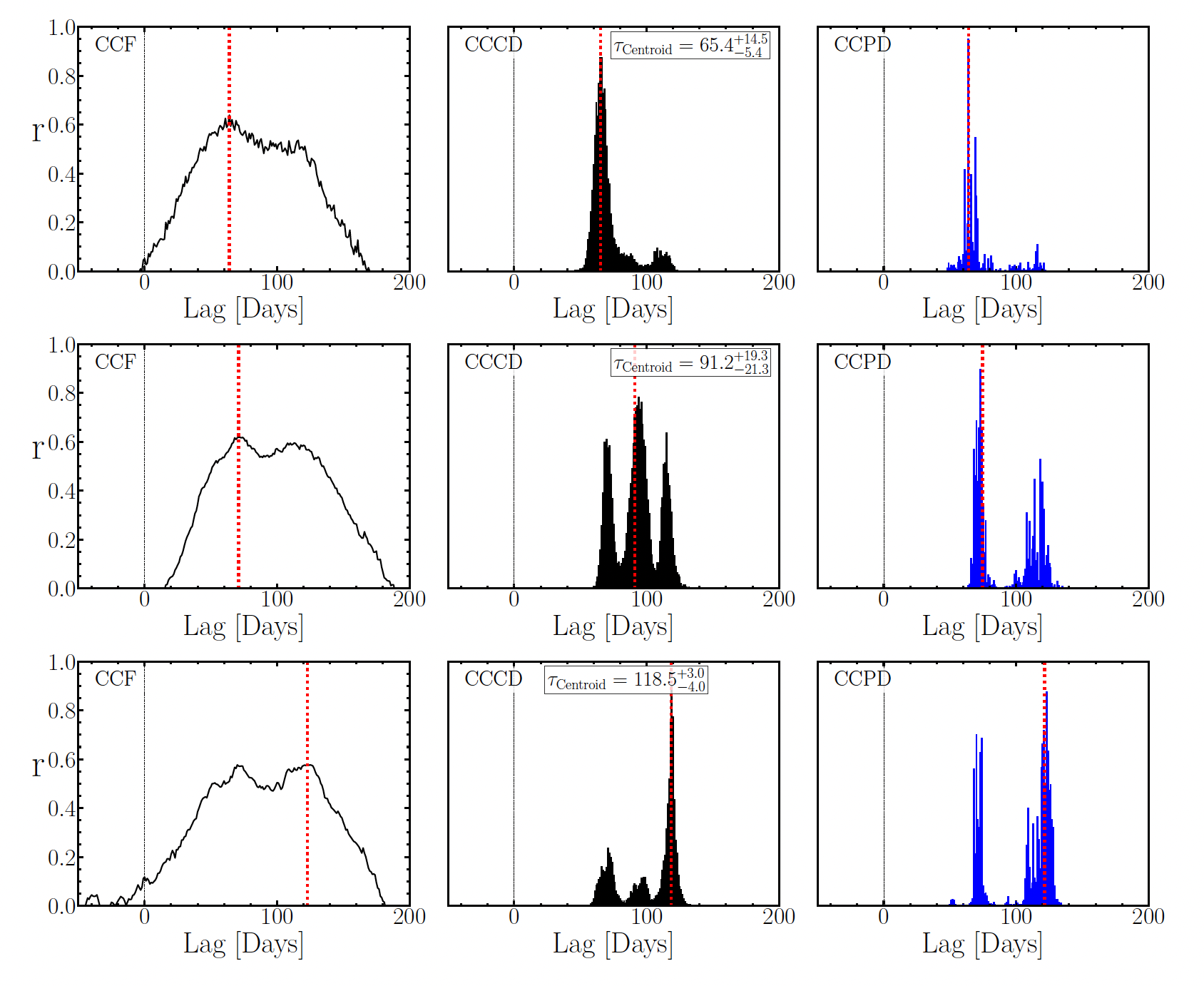}
}
\caption{\label{iccflagtemp} 
As in Fig.~\ref{iccflagyband} for the dust temperature light-curve.}
\end{figure}

Fig.~\ref{templc} shows the light-curves for the hot dust flux in the $H$ band (top panel) and the hot dust temperature (bottom panel). It is clear that the variability pattern of the dust temperature light-curve is very similar to that of the photometric optical light-curves and the spectral accretion disk light-curve (Fig.~\ref{disklc}), whereas that of the dust flux is not. A time-series analysis using PyCCF gives a centroid lags of $\tau = 65^{+15}_{-5}$~days, $91^{+19}_{-21}$~days and $119^{+3}_{-4}$~days for the dust temperature light-curve relative to the $z_s$~band, $g$~band and {\it Swift} UVW2 photometric light-curves, respectively (Fig.~\ref{iccflagtemp}). The CCFs are relatively broad in all three bands, showing three peaks and covering a large range of dust radii ($\sim 40 - 140$~days). This is not unexpected and similar to the situation for the BLR, which has lags from $\sim 2 - 30$~light-days \citep{Storm2-paper2, Storm2-paper5}. But the differences in lags we obtain for the three UV/optical photometric bands are unexpectedly large, given that the $g$~band and $z_s$~band lags relative to the {\it Swift} UVW2 band are only $\sim 3$ and $5$~days, respectively \citep{Storm2-paper1, Storm2-paper7}. These enhanced lag differences seem to be driven by the fact that each of the UV/optical photometric bands prefers a different peak. We used also two additional methods to estimate the differential lags, namely, the Z-transformed Discrete Correlation Function \citep[ZDCF;][]{zdcf} and the Bayesian time lag analysis package MICA \citep{mica2}. These give results consistent with those from PyCCF and are detailed in the Appendix. Our dust temperature response time is similar to the $K$ versus $V$ band flux response time of $\tau = 92\pm9$~light-days obtained by \citet{Kosh14} when Mrk~817 was in an unobscured state.

Overplotted in Fig.~\ref{templc} (top panel) the $z_s$-band light-curve shifted in time by the ICCF lag (brown curve) to the dust flux light-curve shows that it is not as well matched as the temperature variations with notable differences during the periods of changing obscuration as identified by the X-ray monitoring with {\it NICER} of \citet{Storm2-paper3}. In particular, during the period of low obscuration following the X-ray flare and well into the following second period of high obscuration, the dust flux light-curve stays mostly flat at a level between the peaks and troughs exhibited by the optical light-curves. This lack of response is somewhat similar to the reduced response of the {\it Swift} UVW2 to the far-UV {\it HST} light-curve found by \citet{Storm2-paper4} during this period (see their Fig.~6). 

But we cannot exclude that systematics in our absolute flux calibration also play a role in the lack of response of the dust flux light-curve. As for the temperatures, we used the quasi-simultaneous spectral pairs to assess the systematic uncertainties in the spectral flux. We found that their dust fluxes are mostly consistent with each other within $2\sigma$, but a third (7/20 pairs) shows a $3-4\sigma$ difference in the range of \mbox{$\Delta f_{\lambda} \sim (2 - 5) \times 10^{-16}$~erg~s$^{-1}$~cm$^{-2}$~\AA$^{-1}$}. This flux difference could be sufficiently large to mask the true variations. Since the dust temperature depends mainly on the shape of the SED, it is less susceptible to the accuracy of the absolute flux calibration, which could explain why we find a delay for it but not for the flux. Ideally, this experiment should be repeated with space-based near-IR spectroscopy from the {\it James Webb Space Telescope (JWST)}.

\section{The hot dust structure in Mrk~817}

It is of high interest to understand the chemical composition and grain size distribution of the circumnuclear dust in AGN, which can ultimately reveal how cosmic dust forms and evolves in different environments. If we can constrain the location where the hottest dust forms, we can also assess the relationship of the dust to the other components of an AGN, such as the BLR, accretion disk and possible outflows. In the following, we discuss the hot dust structure for Mrk~817 emerging from the STORM~2 campaign.

\subsection{The 'dusty wall'} \label{dustywall}

The campaign of \citet{L19} was the first to present a dust temperature light-curve. However, they were not able to sample it in great detail due to their much lower cadence. Fig.~~\ref{templc} (bottom panel) corroborates their interpretation that the hot dust temperature variations are due to a heating/cooling process introduced by the variability of the UV/optical continuum. This result means that the dust temperature is always comfortably below the sublimation temperature. We considered here only a single grain size, whereas in general the dust will have a grain size distribution. However, it can be shown that this is a reasonable approximation for our limited near-IR wavelength range if it samples a single temperature, since the emission is dominated by the largest and hottest grains. And even for the largest grains, sublimation is effectively instantaneous (on order of a few days) for densities typical of the dusty torus \citep{Baskin18}.

Our result has two implications, as already put forward by \citet{L19} for NGC~5548: (i) the hot dust is most likely dominated by a carbonaceous rather than a silicate composition since the sublimation temperature of carbonaceous dust \citep[$\sim 1800 - 2000$~K;][]{Sal77} is much higher than the average dust temperature of $\sim 1400$~K observed for Mrk~817, whereas silicate dust grains evaporate already at $\sim 1300 - 1500$~K \citep{Lod03}; and (ii) the inner edge of the dusty torus is determined by a process other than sublimation, which makes its location luminosity-invariant. In fact, although Mrk~817 was in an obscured state during the STORM~2 campaign, our dust response time is similar to the $K$ band dust response time of $\tau = 92\pm9$~light-days obtained by \citet{Kosh14} for data taken during the years 2003-2006 when Mrk~817 was in an unobscured state. Such a `dusty wall' was also observed for NGC~4151 using several epochs of both interferometric and reverberation dust radius measurements \citep{Kosh09, Pott10, Schnuelle13}. Although the dust location was seen to change, the radius did not depend on the irradiating luminosity as expected from the radius-luminosity relationship otherwise observed for AGN samples. Instead, a dependence of the dust radius on the historic UV/optical flux $\sim 6$~years prior seemed to be present \citep{Kish13}. 

This paradigm is similar to the picture for protoplanetary disks around young stars, whose dust is composed of large grains most likely due to their settling in the disk midplane \citep{vanBoekel04, Kospal20}. Interferometry finds that the inner dust radius is significantly larger than the dust sublimation radius in many of these disks. This cavity, which is commonly referred to as the `inner hole', is believed to be filled with gaseous disk material that can change its optical thickness and thus influence the location of the puffed-up `wall' or `inner rim' of the dusty, flared and passively illuminated protoplanetary disk \citep{Monnier05, Dull10}. In AGN, the BLR, whose outer radius appears to be dust-limited \citep{L14}, could potentially serve as such a gas reservoir \citep[e.g., as tidally disrupted clumps from the torus as proposed by][]{Wang17}. That the torus hot dust is simply feeding material to the BLR is also supported by the analysis of \citet{Storm2-paper6}. The temperature variations we observe of $\delta T/T \sim 200/1400 \sim 0.1$ (Fig.~\ref{templc}) are similar to the strength of the temperature fluctuations identified in their models on scales of $\sim 100$~light-days that appeared to move slowly ($v \ll c$) inwards. 

Disk-like dust structures in AGN have started to be revealed also by mid-IR interferometric observations achieving an unprecedented high spatial resolution \citep{Isbell22, Isbell23}. The mid-IR emitting dust, which is a larger structure and composed of a mixture of silicates and carbonaceous dust, is expected to connect to the hot 'dusty wall' and form a passive, dusty and flared disk as in protoplanetary disks. The hot dust component of $T \sim 1000$~K sampled by the WISE data (Fig.~\ref{spitzer}) could then be the silicate hot dust, which, if exposed to the same irradiating luminosity as the carbonaceous hot dust, would be located further out (at $\sim 180$~light-days using eq.~\ref{Stefan-Boltz} and our dust response time of $\sim 90$~light-days).

\subsection{The heating source} \label{geometry}

\begin{figure}
\centerline{
\includegraphics[scale=0.7]{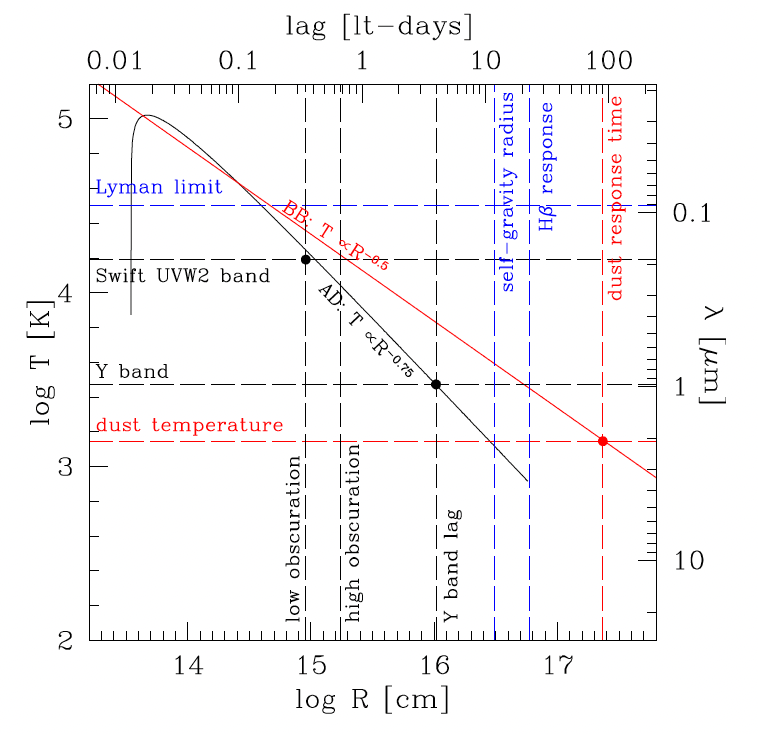}
}
\caption{\label{tempradius} 
The temperature-radius relationship for the hot dust (red solid line) with the observed temperature ($\sim 1400$~K) and response time ($\sim 90$~light-days) indicated by the red horizontal and vertical dashed lines, respectively. The bolometric luminosity illuminating this hot dust is \mbox{$L_{\rm uv} = 6.3 \times 10^{44}$~erg~s$^{-1}$}, corresponding to an accretion rate of $\dot{m} \sim 0.1$. The temperature-radius profile for this accretion disk is shown by the black solid curve. The black horizontal and vertical dashed lines indicate the $9730-9790$~\AA~continuum lag of $\sim 4$~light-days and its central wavelength (labeled $Y$~band), as well as the {\it Swift} UVW2 band rest-frame wavelength and its lag normalizations for the low- and high-obscuration states from \citet{Storm2-paper7} of $\sim 0.4$ and $\sim 0.7$~light-days, respectively. The horizontal blue dashed line marks the Lyman limit of $912$~\AA~beyond which the disk emission is expected to be absorbed in the high-obscuration state. The blue vertical dashed lines indicate the locations of the disk self-gravity radius ($\sim 12$~light-days) and the H$\beta$ BLR lag ($\sim 23$~light-days). The temperatures of the disk blackbodies can be translated to the wavelength of their emission maxima via the Wien displacement law.}
\end{figure}

Following \citet{L19}, we can constrain the luminosity absorbed by the dust, $L_{\rm uv}$, using our estimates of the dust reverberation radius, $R$, and the average dust temperature, $T$, assuming radiative equilibrium as:

\begin{equation}
\label{Stefan-Boltz}
\frac{L_{\rm uv}}{4 \pi R^2} \langle Q^{\rm abs} \rangle = 4 \sigma T^4 \langle Q^{\rm em} \rangle,
\end{equation}

\noindent
where $\langle Q^{\rm abs} \rangle$ and $\langle Q^{\rm em} \rangle$ are the Planck-averaged values of $Q_{\lambda} (a)$ for absorption and emission, respectively, and $\sigma$ is the Stefan-Boltzmann constant. We adopt \mbox{$\langle Q^{\rm em} \rangle=1$}, which is appropriate for relatively large grain sizes of \mbox{$a \ga 0.4~\mu$m} and \mbox{$a \ga 2~\mu$m} for graphite and silicate dust, respectively, emitting as a pure blackbody \citep[see Fig.~8 in][]{L19}. For these grain sizes, $\langle Q^{\rm abs} \rangle$=1 for photons with energies $E\la1$~keV. For the observed average dust temperature ($T \sim 1400$~K) and response radius ($R \sim 90$~light-days) it is $L_{\rm uv} \sim 6.3 \times 10^{44}$~erg~s$^{-1}$. 

This luminosity is similar to the bolometric luminosity of \mbox{$6.7 \times 10^{44}$~erg~s$^{-1}$} estimated for the obscured multi-wavelength SED in Mrk~817 \citep{Storm2-paper1}. This would indicate that soft X-rays are important for heating the dust, as found for NGC~5548 \citep{L19}, and so that the dust 'sees' the obscurer just like the observer does. In this case, however, we expect the torus dust to have a higher temperature in an unobscured state. Using the bolometric luminosity of the unobscured SED and our dust radius, we estimate $T \sim 1650$~K. Such a high dust temperature would support our interpretation that the hottest dust is composed of carbonaceous grains, but this value has not been detected so far. The near-IR spectra of Mrk~817 from the years 2004-2007 revealed hot dust of $T \sim 1400$~K \citep{L11a}, as found during the STORM~2 campaign.  

A second possibility is that the hot dust is insensitive to the soft X-ray obscuration seen by the observer and is heated only by the (unobscured) UV/optical disk emission. Then, the estimated irradiating luminosity corresponds to a disk with an accretion rate of $\dot{m} \sim 0.1$, i.e. a factor of $\sim 2$ lower than the unobscured SED of \citet{Storm2-paper1}. In Fig.~\ref{tempradius}, we plot the temperature-radius (T-R) relationship for a standard disk with $L_{\rm bol} \sim 6.3 \times 10^{44}$~erg~s$^{-1}$, which has the usual form $T \propto R^{-0.75}$. We do not know how far this disk extends, but the $9730-9790$~\AA~continuum lag versus the {\it Swift} UVW2 band of $\sim 4$~light-days (Section~\ref{accdisclag}) puts a useful lower limit. Using the Wien displacement law, we estimate a temperature of $T = 2971$~K for the disk blackbody that has its emission maximum at this wavelength, which makes this lag consistent with the T-R relationship for the $\dot{m} \sim 0.1$ disk. 

Next, we ask which of the {\it Swift} UVW2 band normalizations obtained by \citet{Storm2-paper7} (see Section~\ref{target}) is most consistent with this disk. We find that it is that for the low-obscuration state. This means that the {\it Swift} UVW2 band emission in the high-obscuration state must be dominated by a different emission component, which is either not seen by the dust or is not energetically important for it. The latter case would apply to the BLR DC emission (see also Fig.~\ref{rmsspecdc}) and, intriguingly, the estimated luminosity heating the dust is similar to the (unobscured) bolometric luminosity of the accretion disk assumed by \citet{Storm2-paper10} to explain the UV/optical continuum lags in the high-obscuration state being dominated by the BLR DC emission. Finally, we note that this $\dot{m} \sim 0.1$ disk reaches the self-gravity radius of $\sim 12$~light-days \citep{Lobban22} when the gas in the disk is cool enough to harbour dust of the mean observed temperature. Then, possibly some of the near-IR flux is due to this disk dust component. If this dust is present, a Failed Radiatively Accelerated Dusty Outflow \citep[FRADO;][]{Czerny16, Czerny17} could be launched at these radii and give rise to the H$\beta$ BLR, whose radius of $\sim 23$~light-days lies beyond it. For such an outflow, carbon dust would provide the largest opacity and lead to an inflated disk structure \citep{Baskin18}. The self-gravity radius is also similar to the mean emissivity radius of the BLR DC and \HeII~emission estimated by \citet[][see their Table~1]{Storm2-paper10}, i.e. about half the H$\beta$~BLR radius, and to the \CIV~lag measured by \citet{Storm2-paper5} for the high-obscuration state. Therefore, even a successful outflow might be launched at these radii, as suggested by \citet{Netzer20, Netzer25}.

\subsection{The extended hot dust component and the narrow line region} \label{extdust}

\begin{figure}
\centerline{
\includegraphics[scale=0.2]{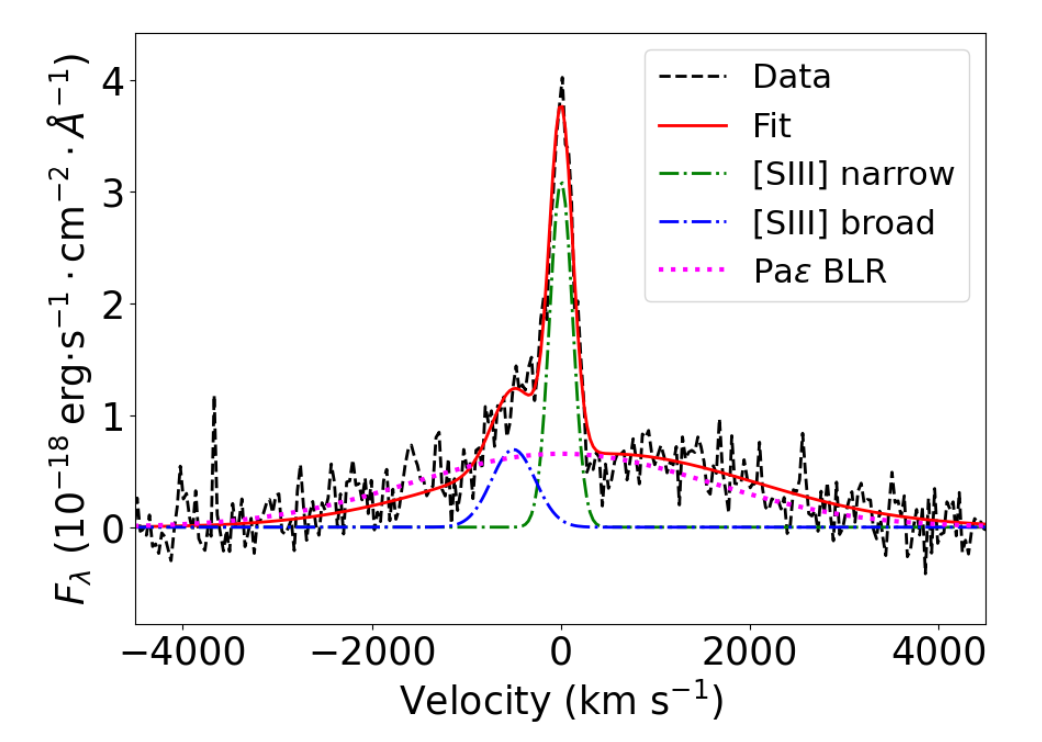}
}
\caption{\label{nifssiiiline} The profile of the \SIII~$\lambda$9531 narrow emission line in the Gemini/NIFS IFU observation was fitted for each individual spaxel with a broad and narrow component. The spatial distribution of these two components is shown in Fig.~\ref{nifs}.}
\end{figure}

The detailed WIYN $H$ band images of \citet{Bentz18} revealed a relatively compact host galaxy for Mrk~817 that harbours a bar and unusually bright near-IR emission associated with the inner bulge. They measured a radial extent for the galaxy disk of $\sim 8\arcsec$ and a bar radius of $\sim 5\farcs6$. The dimensions of the latter are similar to the radius of $\sim 6\arcsec$ we infer for the impressive starburst ring visible on the composite HST/WFC3 image of Mrk~817\footnote{https://hubblesite.org/contents/media/images/2009/25/2577-Image.html}. The residuals of the HST/WFC image of \citet{Bentz09} further show that the bar emerges from an inner ring and then merges into the starburst ring. A putative nuclear starburst in Mrk~817 was observed by \citet{Imanishi04}, who obtained IRTF $L$-band spectra with a $1\farcs6$ wide slit and detected a 3.3~$\mu$m PAH feature on the scale of the bulge. Therefore, it could be that the extended hot dust component we detect as an offset in the Gemini/ARC light-curves is starburst-heated. Interestingly, extended hot dust emission on kpc-scales has been seen in high-redshift quiescent galaxies observed with {\it JWST} \citep{Ji24}. Such hot dust on larger galaxy scales is most likely heated by a young, more spatially extended generation of star formation. 

Since the estimated physical scale for the extended hot dust places it within the narrow-line region (NLR), which has typical dimensions of $\sim 100 - 300$~pc in resolved ground-based observations of nearby Seyfert galaxies, we have further investigated the distribution of the \SIII~$\lambda 9531$~line emission in our NIFS IFU observation. Medium-resolution optical spectroscopy revealed early on a strong blue wing for the \OIII~$\lambda 5007$ narrow emission line \citep{Ilic06}. Such a blue wing is also visible for the \SIII~$\lambda 9531$ line (Fig.~\ref{nifssiiiline}). We then performed a decomposition of the line profile into a broad and narrow component for every spaxel and maps for the corresponding flux distribution and kinematics are presented in Fig.~\ref{nifs} of the Appendix. These maps reveal that the narrow component is resolved and extends on the scales of the host galaxy bulge of $\sim 1\arcsec$. It shows hints of rotation perpendicular to the bar, with the receding part to the south and approaching part to the north. Therefore, the nuclear starburst that heats the dust in the bulge most likely also photoionizes the gas producing the extended \SIII~line emission. 

On the other hand, the outflow, as mapped by the broad component, is unresolved. The broad \SIII~$\lambda 9531$ line profile is blueshifted by $v=-495\pm7$~km~s$^{-1}$ and has a velocity dispersion of $\sigma = 235\pm10$~km~s$^{-1}$. In the Appendix, we estimate an upper limit for the mass outflow rate and kinetic power of $\dot{M}=2.52$\,M$_\odot$\,yr$^{-1}$ and $\dot{E}= 4.92 \times 10^{41}$\,erg\,s$^{-1}$, respectively, with estimated uncertainties that can reach an order of magnitude. The latter nominal value corresponds to $\approx$\,0.04\%\,$L_{\rm bol}$, assuming the total bolometric luminosity of the unobscured SED (Section~\ref{target}), which is one order of magnitude lower than the 0.5\% threshold for significant feedback to the galaxy \citep{Hopkins10}. If this outflow is dusty, e.g. originated as a thermal wind from the obscuring torus, it could be the source of dust and star formation we see for the Mrk~817 galaxy.

\section{Summary and conclusions}

We conducted the first intensive spectroscopic near-IR monitoring campaign on an AGN. Our campaign on Mrk~817 was part of AGN STORM~2 and covered the period of \mbox{2020 Dec 3 - 2022 Aug 2} with a total of 157 cross-dispersed ($\sim 0.7-2.5~\mu$m) spectra, reaching an average cadence of $\sim 3$~days in the first season and $\sim 6$~days in the second season. Our main results for the study of the hot dust can be summarised as follows.

\vspace*{0.2cm}

(i) Our spectral decomposition separates the blackbody dust spectrum from the variable near-IR disk spectrum, but we find that the dust continuum light-curve is relatively flat and does not follow the variability observed for the UV/optical photometric light-curves. On the other hand, the variability of the dust temperature does follow it and we derive time delays of $\tau = 65^{+15}_{-5}$~days, $91^{+19}_{-21}$~days and $119^{+3}_{-4}$~days relative to the $z_s$-band, $g$-band and {\it Swift} UVW2 photometric light-curves, respectively.

(ii) The fact that we can recover a reverberating dust temperature means that the hot dust is always comfortably below the sublimation temperature. Therefore, its composition is most likely dominated by carbonaceous dust since its sublimation temperature \mbox{($\sim 1800 - 2000$~K)} is much higher than the average dust temperature of $T \sim 1400$~K we observe for Mrk~817. Then, since a process other than sublimation sets the dust inner radius, it is most likely luminosity-invariant and so a 'dusty wall'.

(iii) Assuming thermal equilibrium for dust optically thick to the incident radiation, we derive from the average dust temperature and the dust response time the bolometric luminosity of the source heating it, which is $L_{\rm bol} \sim 6 \times 10^{44}$~erg~s$^{-1}$. This luminosity is similar to the bolometric luminosity of the {\it obscured} multi-wavelength SED in Mrk~817, which would imply that soft X-rays are important for the dust heating. Alternatively, the dust is heated by the {\it unobscured} SED of a disk with a low accretion rate ($\dot{m} \sim 0.1$), which can also explain the UV/optical continuum lags in the high-obscuration state as being dominated the BLR diffuse continuum emission \citep{Storm2-paper10}.    

(iv) We detect an extended hot dust component on scales $\ga 140-350$~pc, which from spectra with different slit widths appears as a roughly constant offset between the light-curves. The estimated physical scales place it within the compact bulge of the barred spiral host galaxy and make it also co-spatial with the narrow-line region. For the latter, our Gemini/NIFS IFU observations of the \SIII~$\lambda 9531$ line reveal rotation on the scales of the host galaxy bulge of $\sim 1\arcsec$ and perpendicular to the bar. The extended hot dust could be heated by the nuclear starburst evident on the {\it HST} images of Mrk~817, which then most likely also photoionizes the gas producing the extended \SIII~line emission.

\bibliography{references}{}
\bibliographystyle{aasjournalv7}

\begin{acknowledgments}

The AGN~STORM~2 project began with our successful Cycle 28 Hubble Space Telescope (HST) proposal 16196 \citep{Pet20hst}. Support for HST program GO-16196 was provided by the National Aeronautics and Space Administration (NASA) through a grant from the Space Telescope Science Institute (STScI), which is operated by the Association of Universities for Research in Astronomy (AURA), Inc., under NASA contract NAS5-26555. We are grateful to the dedication of the STScI staff who worked hard to review and implement this program. We particularly thank the Program Coordinator, W. Januszewski, who made sure the intensive monitoring schedule and coordination with other facilities continued successfully. H.L., D.K., J.A.J.M., and M.J.W. were astronomers observing with the Infrared Telescope Facility (IRTF), which is operated by the University of Hawaii under contract 80HQTR24DA010 with NASA. We thank the staff at the IRTF, Gemini Observatory and Apache Point Observatory (APO) for their excellent support with the preparation and efficient execution of the near-IR spectroscopy. The raw data underlying this publication can be downloaded from the Gemini Observatory archive (https://archive.gemini.edu/searchform) by searching for Target:~Mrk~817 and Instrument:~GNIRS, Instrument:~NIFS, and from the IRTF Data Archive (https://irtfweb.ifa.hawaii.edu/research/irtf\_data\_archive.php) by searching for Object:~Mrk~817. This publication makes use of public data products from the Two Micron All Sky Survey (2MASS), which is a joint project of the University of Massachusetts and the Infrared Processing and Analysis Center (IPAC)/California Institute of Technology funded by NASA and the National Science Foundation (NSF), and from the Spitzer Science Heritage Archive held at the NASA/IPAC Infrared Science Archive. Based partly on observations obtained at the Hale Telescope, Palomar Observatory, as part of a collaborative agreement between the Caltech Optical Observatories and the Jet Propulsion Laboratory [operated by Caltech for NASA]. H.L. acknowledges a Daphne Jackson Fellowship sponsored by the Science and Technology Facilities Council (STFC), UK, and financial support from STFC grants ST/P000541/1 and ST/T000244/1. The research by V.G. was carried out at the Jet Propulsion Laboratory, California Institute of Technology, under a contract with the National Aeronautics and Space Administration (grant no. 80NM0018D0004). Research at UC Irvine was supported by NSF grant AST-1907290. J.A.J.M. acknowledges financial support from STFC studentship ST/S505365/1. M.J.W. acknowledges support of an Emeritus Fellowship award from the Leverhulme Trust (EM-2021-064). M.D. and G.F. acknowledge the support from JWST-AR-06419. D.I., A.B.K, and L.\v{C}.P. acknowledge funding provided by the University of Belgrade - Faculty of Mathematics (contract \mbox{451-03-66/2024-03/200104}), Astronomical Observatory Belgrade (contract \mbox{451-03-66/2024-03/200002}), through grants by the Ministry of Education, Science, and Technological Development of the Republic of Serbia. D.I. acknowledges the support of the Alexander von Humboldt Foundation. A.B.K. and L.\v{C}.P. thank the support by Chinese Academy of Sciences President’s International Fellowship Initiative (PIFI) for visiting scientists. C.S.K. is supported by National Science Foundation grants AST-2307385 and 2407206. Y.R.L. acknowledges financial support from the NSFC through grant No. 12273041 and from the Youth Innovation Promotion Association CAS.

\end{acknowledgments}

\vspace{5mm}

\facilities{Gemini(GNIRS), Gemini(NIFS), ARC(TripleSpec), IRTF(SpeX), LCO(optical), Hale(WIRC), HST(STIS)}


\appendix
\restartappendixnumbering

\section{Journal of observations and measurements from the near-IR spectroscopy and photometry}

\startlongtable
\begin{deluxetable*}{lclclrr}
\tabletypesize{\scriptsize}
\tablecolumns{7}
\decimalcolnumbers
\tablecaption{\label{obslog} 
Details of the near-IR spectroscopic observations}
\tablehead{
\colhead{Observation} & \colhead{MJD} & \colhead{Telescope} & \colhead{Airmass} & \colhead{Aperture} & \multicolumn{2}{c}{Continuum S/N} \\
\colhead{Date} & & & & \colhead{(arcsec$^2$)} & \colhead{$H$} & \colhead{$K$}
}
\startdata
2020 Dec  3 & 59186.489913 & ARC          & 1.933 & $1.1 \times 12.0$&  26 &  65 \\
2020 Dec  8 & 59191.498293 & ARC          & 1.733 & $1.1 \times 4.0$ &  78 & 195 \\
2020 Dec 17 & 59200.496114 & ARC          & 1.574 & $1.1 \times 4.0$ &  65 & 146 \\
2020 Dec 21 & 59204.474775 & ARC          & 1.644 & $1.1 \times 4.0$ &  86 & 192 \\
2020 Dec 24 & 59207.441057 & ARC          & 1.848 & $1.1 \times 10.0$&  38 &  64 \\
2020 Dec 26 & 59209.431358 & ARC          & 1.889 & $1.1 \times 7.0$ &  61 & 108 \\
2020 Dec 28 & 59211.453525 & ARC          & 1.657 & $1.1 \times 5.0$ &  60 &  97 \\
2021 Jan  1 & 59215.485787 & ARC          & 1.411 & $1.1 \times 5.0$ &  55 &  93 \\
\enddata

\tablecomments{The columns are: (1) Universal Time (UT) date of observation; (2) Modified Julian Date; (3) telescope; (4) mean airmass; (5) extraction aperture; continuum S/N in the rest-frame wavelength regions of (6) $\lambda = 1.50-1.55~\mu$m, and (7) $\lambda = 2.05-2.10~\mu$m.}

\end{deluxetable*}

\startlongtable
\begin{longrotatetable}
\begin{deluxetable*}{lcl|cccc|cc|l}
\tabletypesize{\scriptsize}
\tablecolumns{10}
\decimalcolnumbers
\tablecaption{\label{lightcurves} 
Details of the near-IR spectroscopic measurements}
\tablehead{
\colhead{Observation} & \colhead{MJD} & \colhead{Telescope} & \colhead{photometric} & \colhead{flux} & \colhead{flux} & \colhead{flux} & \colhead{dust flux} & \colhead{dust flux} & \colhead{$T$} \\
\colhead{Date} &&& \colhead{correction} & \colhead{(9730$-$9790~\AA)} & \colhead{(1.50$-$1.55~$\mu$m)} & \colhead{(2.05$-$2.10~$\mu$m)} & \colhead{(1.50$-$1.55~$\mu$m)} & \colhead{(2.05$-$2.10~$\mu$m)} & \colhead{(K)} \\
&&& \colhead{factor} & \colhead{(erg/s/cm$^2$/\AA)} & \colhead{(erg/s/cm$^2$/\AA)} & \colhead{(erg/s/cm$^2$/\AA)} & \colhead{(erg/s/cm$^2$/\AA)} & \colhead{(erg/s/cm$^2$/\AA)} &
}
\startdata
2020 Dec  3 & 59186.489913 & ARC          & 0.71$\pm$0.07          & (3.559$\pm$0.142)e$-$15 & (3.645$\pm$0.138)e$-$15 & (3.393$\pm$0.052)e$-$15 & (1.757$\pm$0.206)e$-$15 & (2.028$\pm$0.209)e$-$15 & 1497$\pm$32 \\
2020 Dec  8 & 59191.498293 & ARC          & 1.11$\pm$0.04          & (1.863$\pm$0.029)e$-$15 & (2.171$\pm$0.028)e$-$15 & (2.283$\pm$0.012)e$-$15 & (1.675$\pm$0.075)e$-$15 & (2.185$\pm$0.083)e$-$15 & 1377$\pm$11 \\
2020 Dec 17 & 59200.496114 & ARC          & 1.04$\pm$0.05          & (1.970$\pm$0.056)e$-$15 & (2.379$\pm$0.036)e$-$15 & (2.604$\pm$0.018)e$-$15 & (1.751$\pm$0.097)e$-$15 & (2.363$\pm$0.116)e$-$15 & 1358$\pm$11 \\
2020 Dec 21 & 59204.474775 & ARC          & 0.87$\pm$0.03          & (2.307$\pm$0.044)e$-$15 & (2.470$\pm$0.029)e$-$15 & (2.501$\pm$0.013)e$-$15 & (1.436$\pm$0.061)e$-$15 & (1.834$\pm$0.065)e$-$15 & 1396$\pm$12 \\
2020 Dec 24 & 59207.441057 & ARC          & 0.93$^{+0.07}_{-0.08}$ & (2.630$\pm$0.085)e$-$15 & (2.614$\pm$0.069)e$-$15 & (2.492$\pm$0.039)e$-$15 & (1.570$\pm$0.145)e$-$15 & (1.918$\pm$0.161)e$-$15 & 1418$\pm$24 \\
2020 Dec 26 & 59209.431358 & ARC          & 0.70$\pm$0.03          & (3.155$\pm$0.064)e$-$15 & (3.416$\pm$0.056)e$-$15 & (3.418$\pm$0.032)e$-$15 & (1.622$\pm$0.086)e$-$15 & (2.034$\pm$0.090)e$-$15 & 1413$\pm$12 \\
2020 Dec 28 & 59211.453525 & ARC          & 1.26$^{+0.06}_{-0.07}$ & (1.768$\pm$0.051)e$-$15 & (1.918$\pm$0.032)e$-$15 & (1.973$\pm$0.020)e$-$15 & (1.609$\pm$0.099)e$-$15 & (2.102$\pm$0.114)e$-$15 & 1383$\pm$14 \\
2021 Jan  1 & 59215.485787 & ARC          & 0.90$\pm$0.04          & (2.513$\pm$0.064)e$-$15 & (2.445$\pm$0.045)e$-$15 & (2.360$\pm$0.025)e$-$15 & (1.405$\pm$0.085)e$-$15 & (1.745$\pm$0.088)e$-$15 & 1396$\pm$14 \\
\enddata

\tablecomments{The columns are: (1) Universal Time (UT) date of observation; (2) Modified Julian Date; (3) telescope; (4) multiplicative photometric correction factor determined using the narrow emission line~\SIII~$\lambda~9531$ within the \prepspec~routine; uncorrected near-IR spectral fluxes in the rest-frame wavelength regions of (5) $\lambda = 9730-9790$~\AA, (6) $\lambda = 1.50-1.55~\mu$m, and (7) $\lambda = 2.05-2.10~\mu$m; corrected and decomposed near-IR spectral fluxes in the rest-frame wavelength regions of (8) $\lambda = 1.50-1.55~\mu$m and (9) $\lambda = 2.05-2.10~\mu$m, corresponding to the hot dust component, and (10) dust temperature assuming a blackbody emissivity. We note that the values listed in columns (8) and (9) include the large-scale, constant hot dust component identified for the first time in this work.}

\end{deluxetable*}
\end{longrotatetable}

\startlongtable
\begin{longrotatetable}
\begin{deluxetable*}{lc|ccc|ccc|ccc}
\tabletypesize{\scriptsize}
\tablecolumns{11}
\decimalcolnumbers
\tablecaption{\label{wirc} 
Palomar/WIRC near-IR photometry}
\tablehead{
\colhead{Observation} & \colhead{MJD} & \colhead{$J_{\rm ap}$} & \colhead{$J_{\rm PSF}$} & \colhead{PSF} & \colhead{$H_{\rm ap}$} & \colhead{$H_{\rm PSF}$} & \colhead{PSF} & \colhead{$K_{\rm ap}$} & \colhead{$K_{\rm PSF}$} & \colhead{PSF} \\
\colhead{Date} && \colhead{(erg/s/cm$^{-2}$/\AA)} & \colhead{(erg/s/cm$^{-2}$/\AA)} & \colhead{(arcsec)} & \colhead{(erg/s/cm$^{-2}$/\AA)} & \colhead{(erg/s/cm$^{-2}$/\AA)} & \colhead{(arcsec)} & \colhead{(erg/s/cm$^{-2}$/\AA)} & \colhead{(erg/s/cm$^{-2}$/\AA)} & \colhead{(arcsec)}
}
\startdata
2021 Jan 6  & 59220.063022 & (1.91$\pm$0.07)e$-$15 & (2.18$\pm$0.09)e$-$15 & 1.3 & (1.80$\pm$0.09)e$-$15 & (1.86$\pm$0.09)e$-$15 & 1.2 & (2.29$\pm$0.16)e$-$15 & (2.33$\pm$0.14)e$-$15 & 0.9 \\
2021 Feb 3  & 59248.929546 & (2.25$\pm$0.09)e$-$15 & (1.41$\pm$0.06)e$-$15 & 1.0 & (1.88$\pm$0.09)e$-$15 & (1.95$\pm$0.10)e$-$15 & 1.0 & (2.62$\pm$0.15)e$-$15 & (1.92$\pm$0.12)e$-$15 & 1.0 \\
2021 Mar 6  & 59279.848796 & (2.31$\pm$0.09)e$-$15 & (2.20$\pm$0.09)e$-$15 & 0.9 & (1.89$\pm$0.09)e$-$15 & (1.98$\pm$0.10)e$-$15 & 1.0 & (2.62$\pm$0.15)e$-$15 & (2.50$\pm$0.15)e$-$15 & 0.9 \\
2021 May 28 & 59362.000139 & (1.95$\pm$0.09)e$-$15 & (2.13$\pm$0.09)e$-$15 & 1.1 & (2.05$\pm$0.11)e$-$15 & (1.94$\pm$0.10)e$-$15 & 1.1 & (2.65$\pm$0.16)e$-$15 & (2.15$\pm$0.13)e$-$15 & 1.0 \\
2021 Jun 26 & 59391.759444 & (2.02$\pm$0.08)e$-$15 & (2.13$\pm$0.09)e$-$15 & 0.9 & (2.13$\pm$0.11)e$-$15 & (1.97$\pm$0.10)e$-$15 & 0.9 & (2.76$\pm$0.18)e$-$15 & (2.26$\pm$0.17)e$-$15 & 0.7 \\
2021 Aug 9  & 59435.500000 & (1.75$\pm$0.07)e$-$15 & (1.85$\pm$0.07)e$-$15 & 0.8 & (1.79$\pm$0.09)e$-$15 & (1.75$\pm$0.09)e$-$15 & 1.0 & (2.54$\pm$0.12)e$-$15 & (1.88$\pm$0.11)e$-$15 & 1.0 \\
2021 Sep 4  & 59461.713522 & (1.98$\pm$0.08)e$-$15 & (2.13$\pm$0.09)e$-$15 & 1.3 & (2.05$\pm$0.10)e$-$15 & -                     & 1.4   & (2.91$\pm$0.17)e$-$15 & (1.40$\pm$0.08)e$-$15 & 1.3 \\
2022 Feb 9  & 59619.500000 & (1.84$\pm$0.07)e$-$15 & (2.02$\pm$0.08)e$-$15 & 1.3 & (1.86$\pm$0.09)e$-$15 & -                     & 1.3   & (2.45$\pm$0.15)e$-$15 & (2.06$\pm$0.12)e$-$15 & 1.2 \\
\enddata

\tablecomments{The columns are: (1) Universal Time (UT) date of observation; (2) Modified Julian Date; for the $J$ band (3) aperture photometry flux, and (4) PSF fit flux and (5) Point Spread Function (PSF) FWHM for aperture photometry; for the $H$ band (6) aperture photometry flux and (7) PSF fit flux and (8) PSF FWHM for aperture photometry; and for the $K_s$ band (9) aperture photometry flux and (10) PSF fit flux and (11) PSF FWHM for aperture photometry.
\\ $^\dagger$The seeing on this night was very bad and so a wider 10 pixel radius was used for the aperture photometry}

\end{deluxetable*}
\end{longrotatetable}

\section{The NIFS IFU observations}

\begin{figure*} 
\centerline{
\includegraphics[scale=0.32]{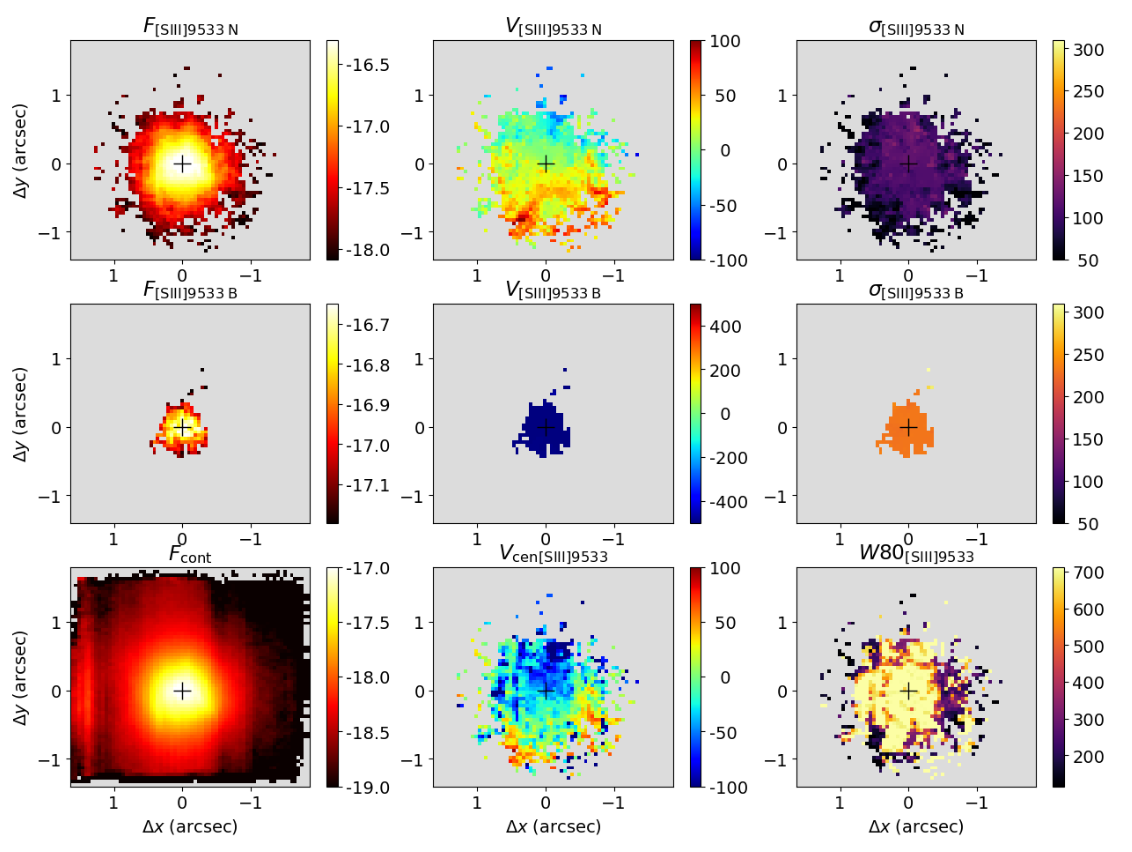}
}
\caption{\label{nifs} 2D maps of the \SIII~$\lambda 9531$ emission line from the IFU observation taken with NIFS on Gemini North on 2022 Jun 4. The flux, velocity shift and velocity dispersion of the narrow (N) and broad (B) components fitted to the line profile as in Fig.~\ref{nifssiiiline} are shown individually in the top two rows. The bottom row shows the 2D map of the continuum flux, the velocity shift of the line center and the velocity width encompassing 80\% of the total line flux (W80).}
\end{figure*}

\begin{figure*} 
\centerline{
\includegraphics[scale=0.32]{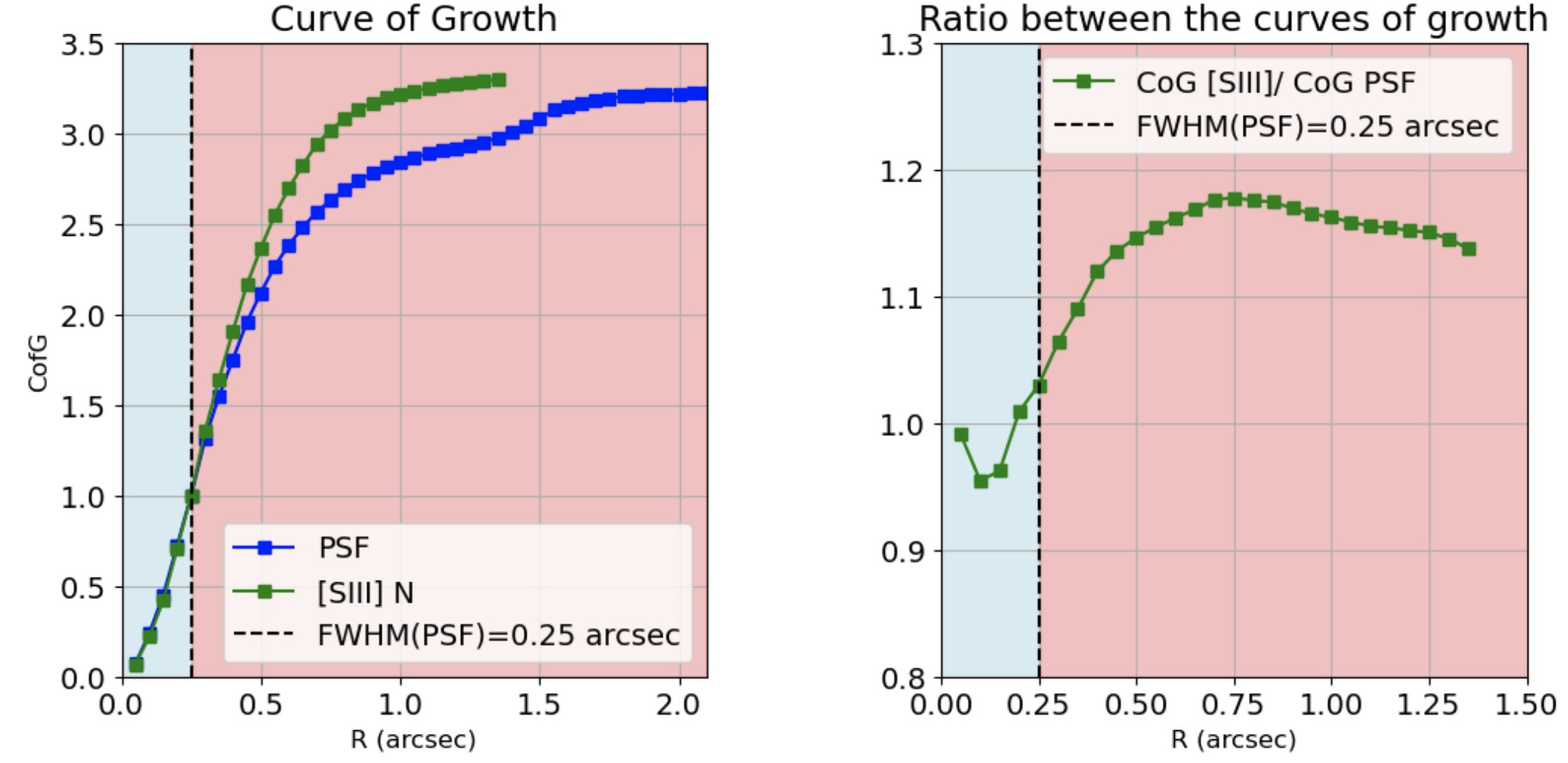}
}
\caption{\label{nifsradius} Curve-of-growth for the unresolved (PSF) flux and the narrow \SIII~$\lambda 9531$ emission line for the NIFS IFU observation (left panel). The ratio between the two curves shows that the \SIII~$\lambda 9531$ emission line region has a radial extent of $\sim 0\farcs8$, with the extended component constituting a maximum of $\sim 20\%$ of the total line flux (right panel).}
\end{figure*}

In Fig.~\ref{nifs}, we present 2D maps of the flux distribution and kinematics in the  \SIII~$\lambda 9531$ emission line measured from the Gemini NIFS data. As shown in Figure \ref{nifssiiiline}, the line profile has two kinematic components: a narrow (N) and a broad (B). The top panel of Figure \ref{nifs} shows that the narrow component is resolved and its velocity field presents a rotation pattern, with blueshifts to the North and redshifts to the South. The middle panel shows that the broad component is completely blueshifted with a velocity close to $-500$\,km\,s$^{-1}$ and is unresolved, because it is confined to within $0\farcs225$ from the nucleus, as the PSF is $0\farcs45$. For a scale of 632\,pc/$\arcsec$, this means that the maximum extent of the outflow is $\approx$ 142\,pc. Within the PSF, the unresolved flux of the broad component is $(9.92\pm0.86) \times 10^{-15}$~erg~s$^{-1}$~cm$^{-2}$.

We present below an estimate for the upper limit of the mass-outflow rate and power of this outflow, assuming it is continuous and mass-conserving. The mass-outflow rate, $\dot{M}$, can be estimated as in \citet{Storm2-paper9} and similarly to \citet{Storchi18}, but for an outflow within a solid angle $\Omega$:

\begin{equation}
\frac{\dot{M}}{\Omega} = f m_p N_e v r^2 = m_p N_H v r,
\end{equation}

\noindent 
where $f$ is the filling factor, $m_p$ is the mass of the proton, $N_e$ is the electronic density, $v$ is the velocity of the broad component and $r$ is the maximum radius of the outflow, as discussed above. For typical NLR values of $f=0.01$, $N_e=1000$\,cm$^{-3}$, using the velocity of the broad component of $v=-495$\,km\,s$^{-1}$ and assuming a maximum radius of $r=142$\,pc, we obtain $\dot{M}/\Omega\approx$\,2.52\,M$_\odot$\,yr$^{-1}$\,sr$^{-1}$. The implied column density of this wind is $N_H = 4.4 \times 10^{21}$~cm$^{-2}$.

The power of the outflow, $\dot{E}$, can be estimated as:

\begin{equation}
\dot{E}/\Omega = \dot{M} (v^2 + \sigma^2),
\end{equation}

\noindent 
where $v=495$\,km\,s$^{-1}$ and $\sigma=235$\,km\,s$^{-1}$. For $\dot{M}=2.52$\,M$_\odot$\,yr$^{-1}$, $\dot{E}=4.92 \times 10^{41}$\,erg\,s$^{-1}$, corresponding to $\approx$\,0.04\%\,$L_{\rm bol}$ and 0.21\%\,$L_{\rm bol}$, assuming the total bolometric luminosity of the unobscured SED of $L_{\rm bol}=1.1 \times 10^{45}$\,erg\,s$^{-1}$ and the $1–1000$~Ryd (13.6 eV to 13.6 keV) ionizing luminosity of the obscured SED of $2.4 \times 10^{44}$~erg~s$^{-1}$, respectively (see Section~\ref{target}). Both the estimates of mass-outflow rate and its power are subject to uncertainties in its density and filling factor, resulting in an uncertainty of about one order of magnitude. Furthermore, given that we rely on the measured \SIII~line intensity, we can only estimate the mass associated with the ionized component. However, the cloud we observe may be optically thick to the ionizing radiation with a mass considerably larger than the one assumed here, due to the mass of the neutral gas.

Only photoionization calculations for a radiation-bound gas remove this uncertainty. Therefore, we carried out dedicated photoionization~\cloudy~calculations for a fully ionized cloud at a distance of $\sim 100$~pc illuminated by the unobscured SED of Mrk~817 and with a hydrogen column density of $N_H = 3.2 \times 10^{21}$~cm$^{-2}$ that show that the \SIII~$\lambda 9531$ emission line luminosity is $\sim 2 \times 10^{42}$~erg~s$^{-1}$. This is about a factor of 100 higher than the observed luminosity of the \SIII~broad line, implying a cloud covering factor of $\sim 1\%$. The total gas mass for this outflowing cloud is $M \sim 3.2 \times 10^4~M_\odot$. Taking the cloud outflow velocity as $v=-495$\,km\,s$^{-1}$ and assuming that the cloud was ejected from locations close to the black hole, the time it took this cloud to arrive at the observed location is $t \sim 200,000$~years. Then, the outflow rate is $\dot{M} = M/t = 0.16 \,M_\odot$ \,yr$^{-1}$ and the kinetic power for the period of 200,000~years is $\dot{E} \sim 3 \times 10^{40}$\,erg\,s$^{-1}$. This estimate is strictly speaking a lower limit since the terminal location of the cloud can be closer in, but this mean value of the mass outflow rate over a period of about 200,000 years is similar to the estimate given by \citet{Netzer25}, which was based on the X-ray observations of the source during the STORM~2 campaign.

\section{Differential lag estimates using ZDCF and MICA} \label{lagmethods}

\begin{figure*} 
\centerline{\fbox{
\includegraphics[scale=0.4]{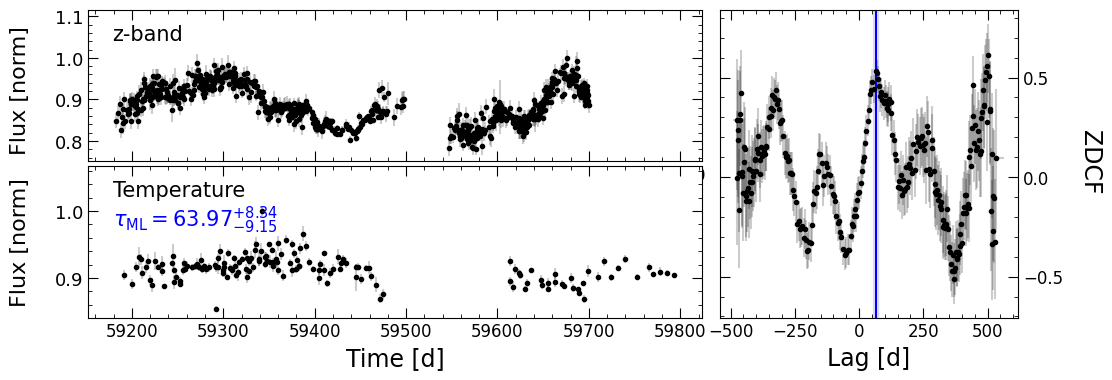}}
}
\centerline{\fbox{
\includegraphics[scale=0.4]{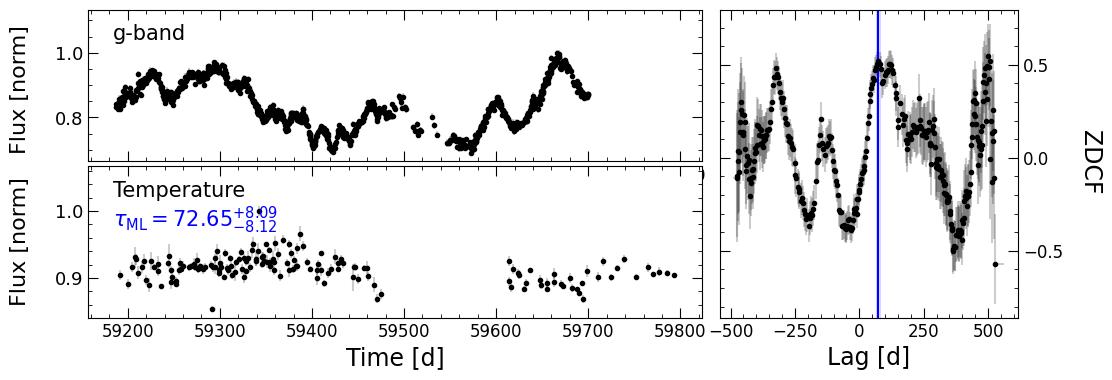}}
}
\caption{\label{zdcf} Differential lag estimates using the ZDCF method for the $z_s$ band versus dust temperature light-curve (top panels) and for the $g$ band versus dust temperature light-curve (bottom panels).}
\end{figure*}

\begin{figure*} 
\centerline{\fbox{
\includegraphics[scale=0.53]{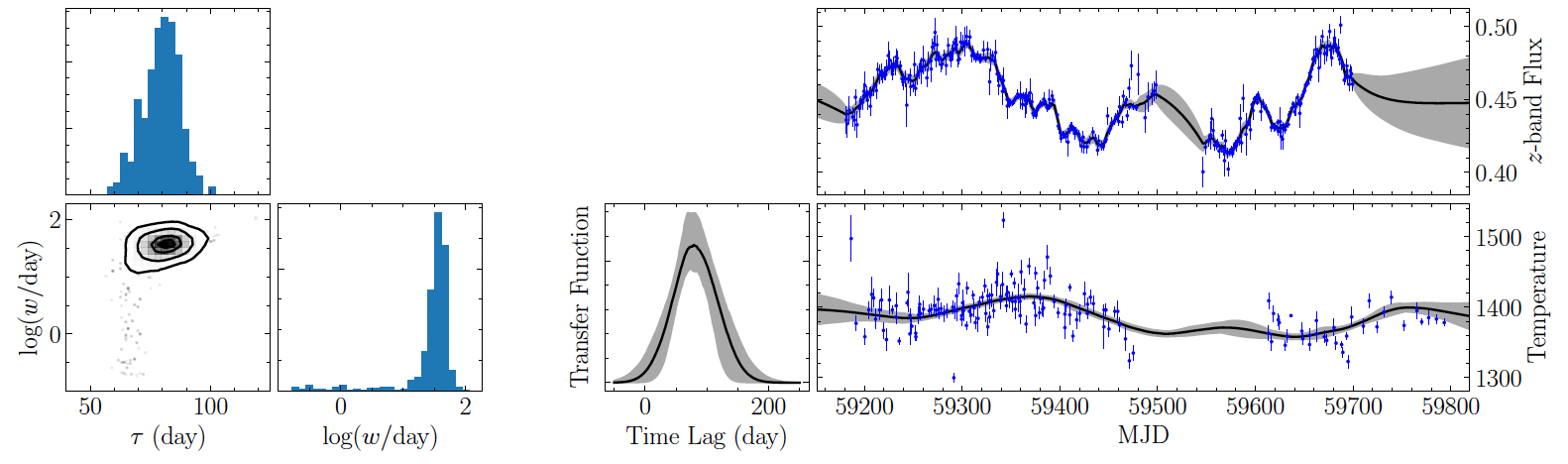}}
}
\centerline{\fbox{
\includegraphics[scale=0.526]{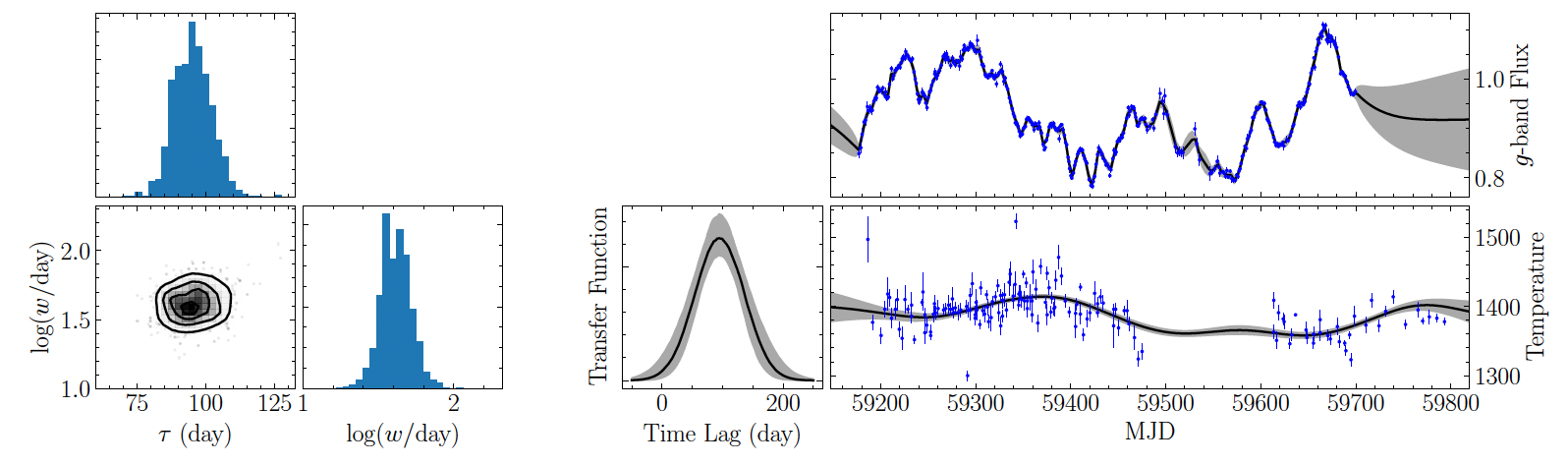}}
}
\caption{\label{mica} Differential lag estimates using MICA with a Gaussian transfer function for the the $z_s$ band versus dust temperature light-curve (top panels) and for the $g$ band versus dust temperature light-curve (bottom panels).}
\end{figure*}

We have applied two additional methods in order to estimate the differential lag between the $g$-band and $z_s$-band photometric light-curves and the dust temperature, namely, the Z-transformed Discrete Correlation Function \citep[ZDCF;][]{zdcf, pyzdcf}\footnote{https://pypetal.readthedocs.io/en/latest/installation.html} and the Bayesian time lag analysis package MICA \citep{mica2}. 

The ZDCF method gives a differential lag between the $z$ band light-curve and the dust temperature light-curve of $\tau = 64^{+8}_{-9}$ days and between the $g$ band light-curve and the dust temperature light-curve of $\tau = 73\pm8$ days. The maximum peak of the ZDCF and its uncertainty were calculated through maximum likelihood methods. These results, which are shown in Fig.~\ref{zdcf}, are consistent within their errors with the results from PyCCF (Section~\ref{revsignal}). Additionally, the presence of symmetric peaks around the detected lags in the ZDCF plots point to reverberation effects within the AGN structure. These effects suggest that light travel times between different emitting regions of the accretion disk and possibly the broader AGN environment contribute to the observed flux variations at different time scales. This pattern of mirrored peaks suggests that responses of the thermal state to changes in accretion dynamics are detectable in our observations.

In MICA, the driving light-curve is modeled with the damped random walk process and the transfer function is parameterized with a basic function. We have explored both a Gaussian and a top-hat transfer function. The parameters of the transfer function and the damped random walk process are sampled by the Markov-chain Monte Carlo (MCMC) technique. From the generated posterior samples, the time lag is assigned as the center of the Gaussian or top-hat and the lower and upper uncertainties are computed from the 16th and 84th percentiles. Since the errors might be underestimated, we have included a free parameter to account for this possibility, which is added in quadrature to the data error. In addition, we rebinned the $g$ and $z$-band light-curves every one day to expedite the running speed. We have checked that such a rebinning does not affect the time lags, since the adopted bin width is far smaller than the time lags. We find that the choice of transfer function has little impact on the results and we get consistent lag values for the two choices. These are $\tau = 80^{+7}_{-9}$~days and $\tau = 83 \pm 5$~days for the $z$ band versus the dust temperature in the case of a Gaussian and top-hat transfer function, respectively. For the $g$ band versus the temperature the results are $\tau = 95^{+7}_{-6}$~days and $\tau = 95^{+5}_{-6}$~days in the case of a Gaussian and top-hat transfer function, respectively. Examples of these results are shown in Fig.~\ref{mica}. The resultant lags are consistent within their errors with those obtained with PyCCF (Section~\ref{revsignal}) and the ZDCF method.

\end{document}